\documentclass[iop]{emulateapj-rtx4}
\usepackage{apjfonts}
\usepackage{array,longtable}

\newcommand{\Msun}{M$_{\odot}$}

\shorttitle{GC Age Uncertainties in IR CMDs}
\shortauthors{Correnti et al.}

\begin{document}


\title{Constraining Globular Cluster Age Uncertainties using the IR Color-Magnitude Diagram\altaffilmark{1}} 


\author{Matteo Correnti\altaffilmark{2}, Mario Gennaro\altaffilmark{2}, Jason S. Kalirai\altaffilmark{2,3}, Thomas M. Brown\altaffilmark{2}, and Annalisa Calamida\altaffilmark{2}}

\altaffiltext{1}{Based on observations with the NASA/ESA {\it Hubble Space Telescope}, obtained at the Space Telescope Science Institute, which is operated by the Association of Universities for Research in Astronomy, Inc., under NASA contract NAS5-26555}

\altaffiltext{2}{Space Telescope Science Institute, 3700 San Martin Drive, Baltimore, MD 21218, USA; correnti, gennaro, jkalirai, tbrown, calamida@stsci.edu}

\altaffiltext{3}{Center for Astrophysical Science, John Hopkins University, Baltimore, MD 21218, USA}

\begin{abstract}
{Globular Clusters (GCs) in the Milky Way are the primary laboratories for establishing the ages of the oldest stellar populations and for measuring the color-magnitude relation of stars. In infrared (IR) color-magnitude diagrams (CMDs), the stellar main sequence (MS) exhibits a ``kink'', due to opacity effects in M dwarfs, such that lower mass and cooler dwarfs become bluer in the IR color baseline. This diagnostic offers a new opportunity to model GC CMDs and to reduce uncertainties on cluster properties (e.g., their derived ages). In this context, we analyzed Hubble Space Telescope Wide Field Camera 3 IR archival observations of four GCs -- 47\,Tuc, M\,4, NGC\,2808, and NGC\,6752 -- for which the data are deep enough to fully sample the low-mass MS, reaching at least $\simeq$ 2 mag below the ``kink''. We derived the fiducial lines for each cluster and compared them with a grid of isochrones over a large range of parameter space, allowing age, metallicity, distance, and reddening to vary within reasonable selected ranges. The derived ages for the four clusters are respectively 11.6, 11.5, 11.2, and 12.1 Gyr and their random uncertainties are $\sigma \sim$ 0.7\,--\,1.1 Gyr. Our results suggest that the near-IR MS ``kink'', combined with the MS turn-off, provides a valuable tool to measure GC ages and offers a promising opportunity to push the absolute age of GCs to sub-Gyr accuracy with the next generation IR telescopes such as the James Webb Space Telescope and the Wide-Field Infrared Survey Telescope.}  
\end{abstract}





\section{Introduction}
\label{s:intro}

Globular Clusters (GCs) are among the oldest objects in the Universe and
accompany most major star formation episodes in galaxies
\citep[e.g.][]{brostr06}. The {\it absolute} age determination of the
population represents one of the most reliable measures of when baryonic
structure formation occurred in the Universe \citep{sper+03} and provides robust
constraints on the physics adopted in stellar evolutionary models
\citep{salwei98,cass+99,vand+08,dott+08}. The {\it relative} age difference
between clusters associated with distinct structural components establishes the
formation and assembly timescales of these parent populations (e.g., the bulge,
halo, and substructure).

Modern derivations of star cluster ages have primarily involved reproducing visible-light color-magnitude diagrams (CMDs) with stellar evolutionary models. However, estimates of absolute GC ages are hampered by uncertainties in other fundamental parameters: in addition to accounting for systematic uncertainties due to different assumptions in stellar models \citep[$\sigma$ = 0.4 Gyr \,--\,][]{chakra02,vall+13a,vall+13b} and the cluster metallicity \citep[$\sigma$ = 0.5 Gyr for a 0.2 dex error \,--\,][]{dott+08}, the largest uncertainty impacting this technique comes from simultaneously ``fitting'' the age at a given distance and reddening \citep[$\sigma$ = 1.5\,--\,2 Gyr \,--\,][]{chab08}. On visible CMDs, the latter quantities are estimated by aligning the luminosity of the horizontal branch and the slope of the unevolved main sequence (MS) to the observations \citep[e.g.,][]{iberen84}. 

Different methods have been proposed to overcome some of these problems: the use of a different clock \citep[white dwarf cooling sequence,][and references therein]{hans+13}, different photometric systems \citep[Str\"{o}mgren bands,][]{grun+98}, or different diagnostics \citep[luminosity function,][]{zocc+00,rich+08}. 
 
\begin{table*}[thb]
\begin{center}
\caption{Description of the datasets}
\begin{tabular}{ccccc}
\hline
\hline
Cluster & Exposure Time & Filter & Program & PI \\
 (1) & (2) & (3) & (4) & (5)\\
\hline
47\,Tuc\footnote{We reported the combined integration time of the {\it Swath} and {\it Stare} fields. For a detailed description of 47\,Tuc observational strategy we refer the reader to \citet{kali+12}.} & 91.8 ks & F110W & 11677 & H. Richer\\
      &  185.6 ks & F160W & \\
M\,4 & 5.2 ks & F110W & 12602 & A. Dieball\\
    & 10.4 ks & F160W &     &           \\
NGC\,2808 & 3 $\times$ 1.4 ks & F110W & 11665 & T. M. Brown\\
         & 3 $\times$ 1.7 ks & F160W &  & \\
NGC\,6752 & 0.7 ks & F110W & 11664 & T. M. Brown\\
         & 0.7 ks & F160W &       &            \\
\hline
\end{tabular}
\tablecomments{Columns (1): Name of the clusters. (2): Total exposure time (kiloseconds). (3): Filter. (4): Program identification number. (5): Principal Investigator.}
\label{t:data}
\end{center}
\end{table*} 

\citet{bono+10} introduced a new tool to measure accurate star cluster parameters, based on high-precision infrared (IR) CMDs. In fact, the MS in a pure IR CMD exhibits a ``kink'' (hereafter we refer to it simply as kink) at $\sim$ 0.5 \Msun, i.e. turns towards bluer colors \citep{pulo+98,pulo+99,zocc+00,cala+09,sara+09,kali+12,milo+12,milo+14,mone+15}. This feature arises from a redistribution of the emerging stellar flux due to a change in opacity caused by the collision-induced absorption of H$_2$ \citep[][and references therein]{lins69,saum+94}. The theoretical predictions of temperature and luminosity of the kink are minimally affected by uncertainties in the treatment of convection (i.e., by mixing length theory calibration), as the convective motions are nearly adiabatic at these masses \citep{saumar08}. The luminosity of the kink and the shape of the bending are dependent on metallicity, but are independent of age beyond $\sim$ 1 Gyr. Therefore, the degeneracy between these two parameters can be partially broken. Equally as important, the shape of the color-magnitude relation in the IR CMD, in particular its inversion at magnitude fainter than the kink, provides a new opportunity to simultaneously constrain the distance and reddening accurately. Therefore, this new diagnostic provides a large lever arm to decrease the uncertainties involved in deriving GC age estimates. 
 
To test the power of this feature to establish accurate GC properties, we
searched the MAST STScI-archive for public Wide Field Camera 3 (WFC3) IR data
sampling a set of Milky Way GCs. There are a dozen clusters for which such data
are available, but, only for four GCs, namely 47\,Tuc, M\,4, NGC\,2808, and
NGC\,6752, the data are deep enough to reach $\simeq$\,2 mag below the kink. 
This depth is required to fully sample the bending of the MS at low masses. The
four GCs provide a good sample to test the predictive power of the MS kink
thoroughly, as they span a significative range of metallicity ([Fe/H] $\simeq$
-0.7\,--\,-1.5 dex), distance ($d{\odot} \simeq$ 2.1\,--\,9.6 kpc) and reddening
(E\,(B-V) $\simeq$ 0.04\,--\,0.39 dex).

Using an ad-hoc fitting method, we derived a fiducial line in the CMD for each GC.  These fiducial lines represent the loci where we expect the single stars (i.e., excluding binaries and higher multiples) of the cluster to lie if they were observed without errors. We compared these fiducials with a grid of isochrones and derived the joint posterior probability density function (PDF) for the four parameters, age, metallicity, distance and reddening. This, in turn, allowed us to estimate the best fitting parameters, quantify the correlations among them and derive the uncertainties of the individual parameters. In this paper, we focus on the derived GC ages, by marginalizing the joint PDF over the remaining three parameters.  

Our analysis demonstrates the importance of the near-IR MS kink as an age diagnostic, and illustrates how this feature can provide the first opportunity to push the absolute ages of GCs to sub-Gyr precision.  

The remainder of the paper is organized as follows: we present the data set and reduction in Section~2 and the CMDs of the four GCs in Section~3.  We also describe the method adopted to measure the fiducial lines for each of the GCs.  In Section~4 we derive the best-fit isochrones and the probability distribution functions, thereby obtaining an estimate of the age and uncertainty for each GC. In Section~5 we discuss the results from our study.

\section{Sample and Data Reduction}
\label{s:reduction}
Our sample consists of four clusters: 47\,Tuc (GO-11677, PI: H. \ Richer), M\,4 (GO-12602, PI: A. \ Dieball), NGC\,2808 (GO-11665, PI: T. M. \ Brown), and NGC\,6752 (GO-11664, PI: T. M. \ Brown). With the exception of 47\,Tuc, for which the photometry has been derived in \citet{kali+12}, we downloaded from the MAST STScI-archive the images for the other three clusters. A description of these datasets is provided in Table~\ref{t:data}.
	 
To reduce the images of M\,4, NGC\,2808 and NGC\,6752 we followed the same
method adopted for 47\,Tuc, described in detail in \citet{kali+12}. Briefly, we
started from the \emph{flt} files provided by the HST pipeline, which constitute
the bias-corrected, dark-subtracted and flat-fielded images. We generated
distortion-free images using MultiDrizzle \citep{fruchook02} and we calculated  
the transformation between the individual drizzled images in each filter,
linking them to a reference frame (i.e., the first exposure). Through these
transformations we obtain an alignment of the individual images to better than
0.02\,--\,0.04 pixels. After flagging and rejecting bad pixels and cosmic rays
from the input images, we created a final image for each filter, combining the
input undistorted and aligned frames. The NGC\,2808 and NGC\,6752 final images were
slightly supersampled to a pixel scale of 0\farcs09 pixel$^{-1}$ to mitigate the
severe undersampling of these data. A square kernel was used in the final image
generation and the pixfrac was kept near unity. M\,4 has a significative number
of overlapping images, so the final images were supersampled from the native
resolution of 0\farcs13 pixel$^{-1}$ to 0\farcs06 pixel$^{-1}$, with a Gaussian
kernel with pixfrac = 0.70. The FWHM on this image is $\sim$ 2.5 pixels.  

To perform the stellar photometry in M\,4, we used the stand-alone versions
of the DAOPHOT-II and ALLSTAR point spread function (PSF) fitting programs
\citep{stet87,stet94} on the stacked images. To obtain the final catalog we
first performed aperture photometry on all the sources that are at least
3$\sigma$ above the local sky.  We then derived a PSF from $\sim$ 1000 bright
isolated stars in the field, and applied this PSF to all of the
sources detected through the aperture photometry. We retained in the final
catalogs only the sources that were iteratively matched between the two images
and we cleaned them eliminating background galaxies and spurious detections by
means of $\chi^2$ and sharpness cuts from the PSF fitting. For NGC\,2808 and NGC\,6752, the data were too undersampled to accurately derive a PSF from the frame with DAOPHOT II. The photometry of all sources was therefore measured using an aperture with R = 2.5 pixels. 

Photometric calibration has been performed using a sample of bright isolated
stars to  transform the instrumental magnitudes to a fixed aperture of 3 pixels.
We  then transformed the magnitudes into the VEGAMAG system by adopting the 
relevant synthetic zero points for the WFC3/IR filters. For NGC 2808, the
final photometric source catalog is built by combining similar observations  in
three separate fields of the cluster. We corrected  each of these fields for
differential reddening prior to merging.

\begin{figure*}[thp]
\hspace*{0.5cm}
\includegraphics[scale=0.6,angle=270]{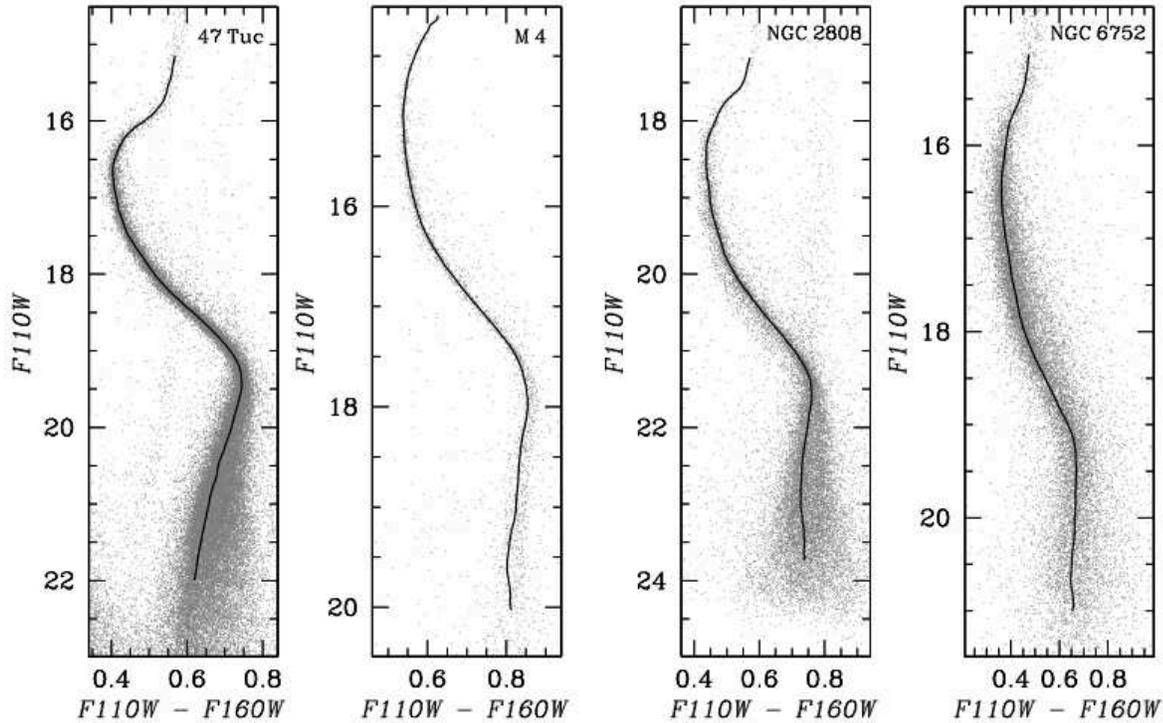}
\caption{$F110W-F160W$ vs $F110W$ CMDs for the four clusters in our sample (from left to right panel: 47\,Tuc, M\,4, NGC\,2808, and NGC\,6752), with superimposed fiducial lines (black lines) that were derived as described in Sect.~\ref{s:cmd}. In contrast to simply deriving the mean color in fixed magnitude bin, the adopted approach avoids the possibility that the shape of the fiducial is altered by the presence of binaries as well as by multiple main sequences, in particular in the lower part of the MS below the kink.}
\label{f:cmd}
\end{figure*}

\section{Color-Magnitude Diagrams and Fiducial lines}
\label{s:cmd}
Fig.~\ref{f:cmd} shows the $F110W-F160W$ vs $F110W$ CMDs for the four clusters (from left to right; 47\,Tuc, M\,4, NGC\,2808, and NGC\,6752). The derived CMDs exhibit very well defined sequences, from the sub giant branch to the low-mass MS well below the kink, which is clearly revealed in all four GCs. NGC\,6752 has the shallowest photometry in our sample, but we are still able to trace the MS at least $\sim$ 2 mag below the kink which allows us to adequately sample its shape and bending at low masses. With the exception of M\,4, all of the cluster CMDs also sample the lower portion of the red giant branch (RGB), which offers additional leverage in simultaneously fitting models to different evolutionary stages.

The first step to obtain accurate GC parameters is to derive a fiducial
line for each of the CMDs. For this, we considered the data in the $F160W$ vs.
$F110W - F160W$ CMD. We used a kernel-density-estimation (KDE) to estimate the
2d PDF underlying the data. The shape of the PDF derived from the KDE can be
thought of as a mountain range whose ridge represents the locus of the true 
underlying cluster isochrone for single stars, while the width of the range is
due mostly to the measurement errors.  Binaries and multiple populations
also play a role in the width. We proceeded by finding the PDF global maximum, which
becomes the starting point of the ridge line search. By moving along the
direction of minimum gradient, we traced the ridge line. Each move was done by
proposing a set of moves of fixed length within an opening cone. The move that
caused the smallest drop in altitude (i.e. in PDF value) was the selected one.
The length of the chosen step is related to the smoothness of the KDE PDF. For
more populous CMDs, the step size can be chosen to be smaller and  the resulting
ridge (fiducial) line is more detailed.  The walk along the ridge line was
performed in two separate iterations. Starting from the same highest peak of
the mountain ridge, one of these was a move towards brighter magnitudes and the 
other was a move towards fainter magnitudes.  Once this ridge line was traced, we ensured the smoothness of the ridge line by computing the corresponding bezier curve and by using this smooth approximation instead of the previously found line.

While the direction of minimum gradient traces the underlying ridge line of the
population, at each step the local direction of maximum gradient represents the
direction along which the data are scattered due to measurement errors. To
estimate the actual values for the fiducial lines and their uncertainty, we
adopted a bootstrap approach. We drew one thousand samples from each cluster
catalog, using the same total number of stars, and allowing for repetitions. We
then constructed a fiducial for each bootstrap catalog using the method outlined
above. The next step was to divide the ensemble of one thousand fiducial
realizations, each with about 200 points, into 0.05 mag bins in $F160W$. This
bin size was chosen to ensure that each bin contains $>= 500$ points, thus
minimizing Poisson noise and avoiding large bulk color changes across the bin
which would otherwise artificially increase the uncertainty.  We took the bin
center as the $F160W$ fiducial value and half the bin size as its error. To
estimate the fiducial values and their errors for the $F110W - F160W$ color, we
took the mean and the standard deviation over the ensemble of fiducials, using
the points that fell within each magnitude bin.

This approach offers two main advantages when compared to simply deriving the mean color in fixed magnitude bins. First, by starting from the mode of the distribution, and by moving along the most prominent range, this approach is not prone to biases due to the presence of binaries which form a mountain range that runs parallel to the main one. It also avoids the possibility that the fiducial shape is altered by the presence of multiple main sequences and ensures that the most prominent one is identified. This is particularly important in the lower part of the MS below the kink, where the separation between the different main sequences is more evident in the IR CMDs. In fact, as \citet{milo+12,milo+14} demonstrated, in NGC\,2808 and M\,4 the MS is clearly separated in two components below the kink, due to the different chemical composition of the two sub-populations. \citet{milo+12,milo+14} suggest that the bluer MS represents a first stellar generation, having primordial helium and high oxygen, while the redder MS is associated with a second generation consisting of stars enhanced in helium, nitrogen, sodium and depleted in oxygen.

\begin{figure}[thp]
\begin{center}
\includegraphics[width=1\columnwidth]{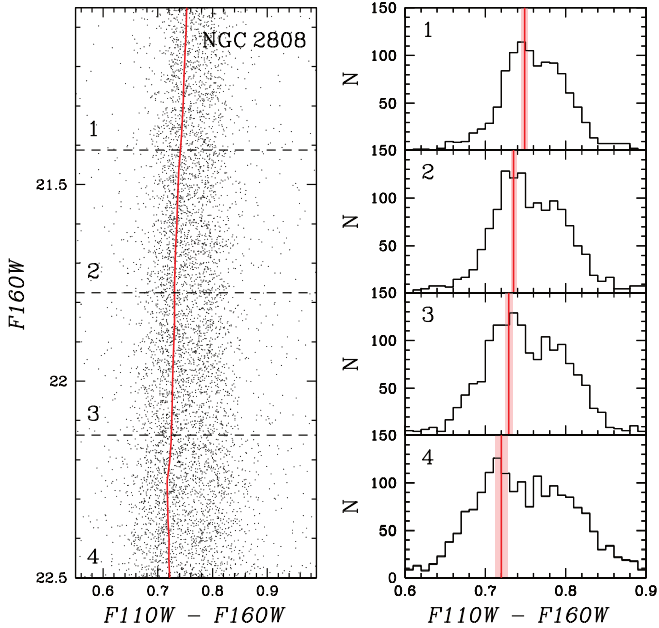}
\caption{Left panel: zoom-in of NGC\,2808 CMD in the region below the kink where the MS split is more evident. The fiducial line, derived as described in Sect.~\ref{s:cmd}, is superimposed as a red curve. Right panel: color distribution of stars in the four magnitude intervals indicated by the dashed lines in the left panel. The mean fiducial color (red line) and its mean error (light red region) in each specific bin is also shown.}
\label{f:split1}
\end{center}
\end{figure}
To verify that our method is able to accurately trace the most prominent population and to demonstrate its strength, we followed the same approach adopted by \citet{milo+12} to identify the MS splitting below the kink. The left panel of Fig.~\ref{f:split1} shows a zoom-in of the NGC~2808 CMD below the kink, where the MS split is more evident. The derived fiducial line is superimposed in red. The color distribution of stars in the four magnitude intervals, over the range $21.05 < F160W < 22.5$ mag, are shown in the right panels.  The mean fiducial color (red line) and its mean error (light red region) in the specific bin is also shown. The distribution is clearly bimodal, with an overall shape similar to the one found by \citet{milo+12}. The different number of stars in each bin is caused by different choices in the selection of the sample, and the mean color of the fiducial line is almost coincident with the peak of the distribution in each bin. Hence, Fig.~\ref{f:split1} confirms that the adopted method successfully traces the most prominent population of the cluster when more than one main sequence is present.  

\begin{figure}[thp]
\begin{center}
\includegraphics[width=1\columnwidth]{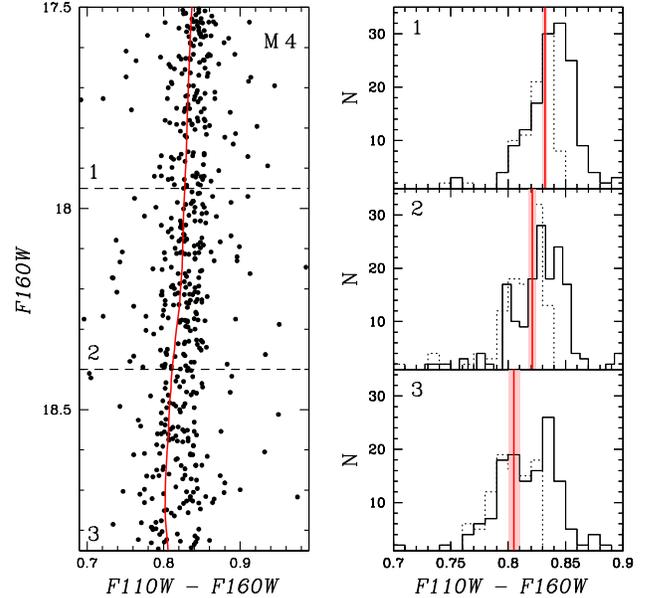}
\caption{Same as Fig.~\ref{f:split1} but for the cluster M\,4. In the case of M\,4, the primordial population is not the most populous one, therefore we have modified the fiducial line computation mechanism to follow the secondary ridge of the density distribution and not the primary, as shown in the right panel (see text for details). The dotted histograms represent the color distribution of stars in the three magnitude intervals after applying the magnitude-color cut described in Sect.~\ref{s:cmd}.}
\label{f:split2}
\end{center}
\end{figure}

To derive the ``genuine'' GC parameters, it is important to trace the primordial ``first generation'' population that has the original cluster chemical abundance. In NGC\,2808, the first stellar generation is also the most populated one, but this is not the case in M\,4. As pointed out by \citet{milo+14} and shown in the right panel of Fig.~\ref{f:split2}, the blue sequence in M\,4 is less populated than the red one. Hence, for this cluster, our method does not properly derive the fiducial line of the cluster below the kink. To overcome this issue, we adopted the following empirical approach: we applied a magnitude-color cut in the CMD to select the stars fainter than $F160W \sim 17$ mag where the split between the two sequences begins.  The cut is made to select stars that are redder than the quantity $(F110W - F160W)_{fiducial} + \sigma$, where $(F110W - F160W)_{fiducial}$ is the color of the fiducial line that traces the most prominent population and $\sigma$ is its error. We then removed these objects from the M\,4 catalog and re-derived the fiducial line and its error on the new dataset.  We iterated this procedure by applying different magnitude cuts through a $\pm 0.25$ mag step with respect to the default quoted above, and verified that the derived fiducial lines are consistent with each other within the errors and that the shape is not altered by the different selection choices. 
\begin{table*}[thb]
\begin{center}
\caption{GC prior parameters and ranges}
\begin{tabular}{cccccccccc}
\hline
\hline
Cluster & Age$_{range}$ & [Fe/H] & $\Delta$ [Fe/H] & ref. & $(m-M)_0$ & $\Delta$ $(m-M)_0$ & E\,(B-V) & E\,(B-V)$_{range}$ &ref.\\
 (1) & (2) & (3) & (4) & (5) & (6) & (7) & (8) & (9) & (10)\\
\hline
47\,Tuc   & 9.0\,--\,15.0  & -0.77  & $\pm$ 0.20 & 1 & 13.24 &  $\pm$ 0.25 & 0.04 & 0.00\,--\,0.20 & 1\\
M\,4      & 9.0\,--\,15.0  & -1.12  & $\pm$ 0.20 & 2 & 11.28 &  $\pm$ 0.25 & 0.37 & 0.22\,--\,0.52 & 2\\
NGC\,2808 & 9.0\,--\,15.0  & -1.22 & $\pm$ 0.20 & 3 & 14.91 &  $\pm$ 0.25 & 0.19 & 0.04\,--\,0.34 & 1\\
NGC\,6752 &  9.0\,--\,15.0 & -1.55  & $\pm$ 0.20 & 4 & 13.01 &  $\pm$ 0.25 & 0.04 & 0.00\,--\,0.19 & 1\\
\hline
\end{tabular}
\tablecomments{Columns (1): Name of the clusters. (2): Age interval (Gyr). (3): Metallicity reference value (dex). (4): Metallicity interval (dex). (5): Reference for the metallicity value ((1) \citet[][and references therein]{carr+09b}; (2) \citet{mari+11}; (3) \citet{mari+14}; (4) \citet[][and references therein]{carr+09b}. (6): Distance modulus reference value (mag). (7): Distance modulus interval (mag). (8): Reddening reference value (mag). (9): Reddening interval (mag). (10) Reference for distance modulus and reddening values. ((1) \citet[][updated as in 2010]{harr06}; (2) \citet{hend+12}).}
\label{t:GC_prior}
\end{center}
\end{table*} 
The resulting color histograms (dotted line) for each magnitude bin are reported in the right panel of Fig.~\ref{f:split2}, together with the mean fiducial color (red line) and its mean error (light red region). With the exception of the central magnitude interval, for which the mean color of the fiducial is still too red compared to the color of the blue peak of the first stellar generation, the newly derived fiducial line reproduces quite well the observed distribution. However, as we will see in Sect.~\ref{s:iso}, this small discrepancy does not affect the global fitting and the resulting best-fit parameters for M\,4. Due to the different procedure applied for this cluster and the low number of stars below the kink, the uncertainties derived in Sect.~\ref{s:prob} could be slightly underestimated.

Tracing the primordial MS below the kink is an important factor in deriving the correct cluster metallicity and therefore a precise age estimate. In fact, the degree of the MS bending is strictly related to the cluster metallicity, due to the different opacities in stars. The net effect is that more metal-rich clusters will exhibit a MS bending that is more accentuated bluewards. This effect is clearly visible in the fiducial lines shown in Fig.~\ref{f:cmd}; as expected, the more metal-rich GC of our sample (i.e., 47\,Tuc, [Fe/H] $\simeq$ -0.75 dex, left panel of Fig.~\ref{f:cmd}) exhibits a fiducial line which is clearly more tilted bluewards than the others. Moving towards lower metallicity (from the left to the right CMDs in Fig.~\ref{f:cmd}), the fiducial lines bending shifts progressively redwards, until in the most metal-poor GC of our sample (i.e., NGC\,6752, [Fe/H] $\simeq$ -1.5 dex, right panel of Fig.~\ref{f:cmd}) there is essentially no bending and the sequence is vertical.

\section{Constraints on the GC parameters} 
  
\subsection{Isochrone fitting: deriving ages}
\label{s:iso}
To derive the age, metallicity, distance and reddening for the four GCs, we compared the fiducial lines (obtained as described in Sect.~\ref{s:cmd}) with a set of stellar models from the Victoria-Regina stellar evolution code \citep{vand+14}.  The models were transformed to the WFC3/IR filter system by performing synthetic photometry using the MARCS library of stellar spectra \citep{gust+08} and the most updated WFC3/IR throughputs and zero points. Extinction was taken into account by applying the \cite{fitz99} extinction law to the spectra before integrating them under the throughput curves, using the appropriate $R_V$ value for each cluster. 

We constructed a grid of isochrones over a large range of parameter space that allowed age, metallicity, distance and reddening to vary over reasonable ranges. The reference values and the adopted intervals are reported in Table~\ref{t:GC_prior}. We use a fixed age range from 9 to 15 Gyr for all GCs, whereas for the metallicity, we used an interval of $\pm$ 0.20 dex with respect to literature values derived from spectroscopic studies for each cluster. For the distance modulus, we adopted an interval of $\pm$ 0.25 mag.  For the reddening, we used $E\,(B-V)=0$ as a lower limit and $E\,(B-V)_{ref} +0.15$ mag as an upper limit (with the exception of M\,4, for which we adopted an interval of $\pm$ 0.15 mag from the reference value). When necessary, we iteratively readjusted the parameter range in order to center the derived best-fit results in the new interval. The derived grid is obtained by adopting steps of 0.1 Gyr for the age, 0.01 dex for the metallicity and 0.01 mag for distance and reddening. We assume that, within the allowed ranges, the prior probability distribution of the parameters is uniform.

In our analysis, the only a priori fixed parameters are the [$\alpha$/Fe] ratio, the extinction coefficient $R_V$ and the helium abundance. For the [$\alpha$/Fe] ratio we adopted the value [$\alpha$/Fe] = +0.4, as suggested from spectroscopic studies of these clusters.  For the extinction coefficient, $R_V$, we adopted the value $R_V = 3.1$ for all GCs except for M\,4, for which we adopted a higher value of $R_V = 3.6$ \citep{hend+12}. The original helium abundance in the stellar models was obtained using Eq.~1 and Eq.~2 from \citet{genn+10}, after the ratio between [Fe/H] and [M/H] was taken into account using Eq.~3 from \citet{sala+93}. We assumed the solar mixture from \citet{aspl+09}, a primordial helium abundance of $Y_P = 0.2485$ from \citet{izot+07} and \citet{peim+07}, and a ratio of $\Delta Y / \Delta Z$ = 1.5. 

To obtain the best-fit isochrone for each cluster, we compared the fiducial lines with the isochrone grid, deriving for each isochrone the posterior PDF which, due to the choice of uniform priors for our parameters, is proportional to the likelihood $\mathcal{L}$. The latter is calculated using the following equation:
\begin{equation}
\mathcal{L} \simeq  \exp\,(-\frac{1}{2} \chi^{2})
\label{e:like}
\end{equation}
where the term $\chi^2$ in Eq.~\ref{e:like} is defined as:
\begin{equation}
\chi^2 = \sum_{i=1}^{N}\frac{(\Delta col_{i})^{2}}{\sigma^{2}_{i}}
\label{e:chi}
\end{equation}
where $\Delta col$ is the difference in color between the isochrones and the fiducial (i.e. $(F110W - F160W)_{iso} - (F110W - F160W)_{fiducial}$), calculated at each point in the fiducial line. $\sigma$ is the error associated with the fiducial color points, derived as described in Sect.~\ref{s:cmd}. The best-fit isochrone is the one that maximizes the joint PDF for the four parameters. 

\begin{figure*}[thp]
\hspace*{0.5cm}
\includegraphics[scale=0.6,angle=270]{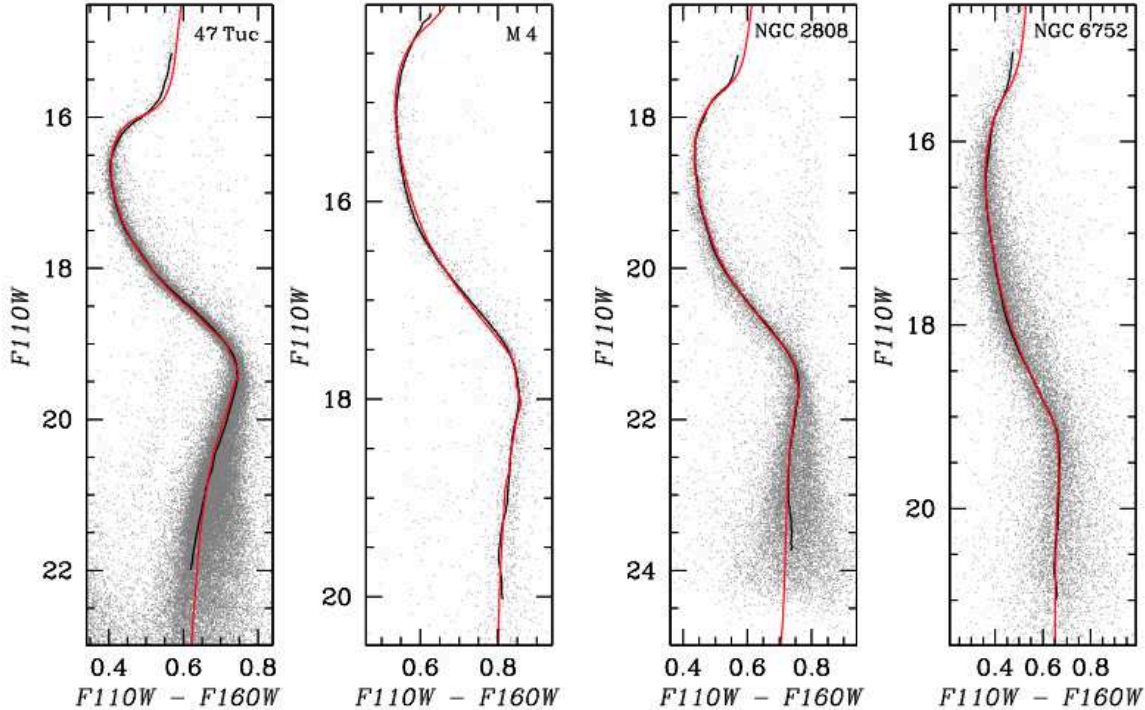}
\caption{$F110W-F160W$ vs $F110W$ CMDs for the four clusters in our sample (from left to right panel: 47\,Tuc, M\,4, NGC\,2808, and NGC\,6752), with superimposed fiducial lines (black lines) and the best-fit isochrones derived as described in Sect.~\ref{s:iso}. Best-fit parameters are reported in Table~\ref{t:GC_param}.}
\label{f:cmd_iso}
\end{figure*}

\begin{figure*}[thp]
\centerline{
\includegraphics[scale=0.45]{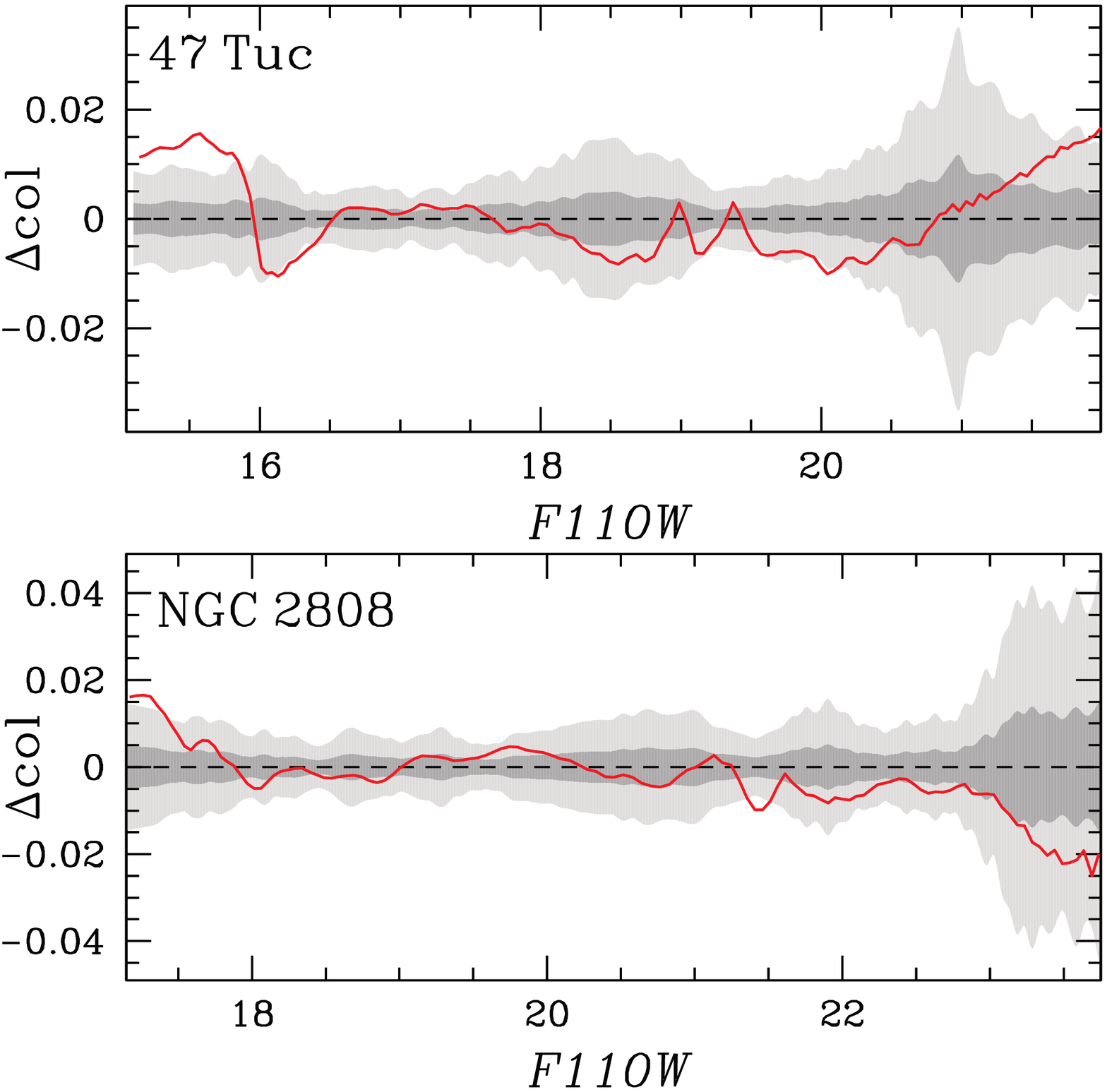}
\includegraphics[scale=0.45]{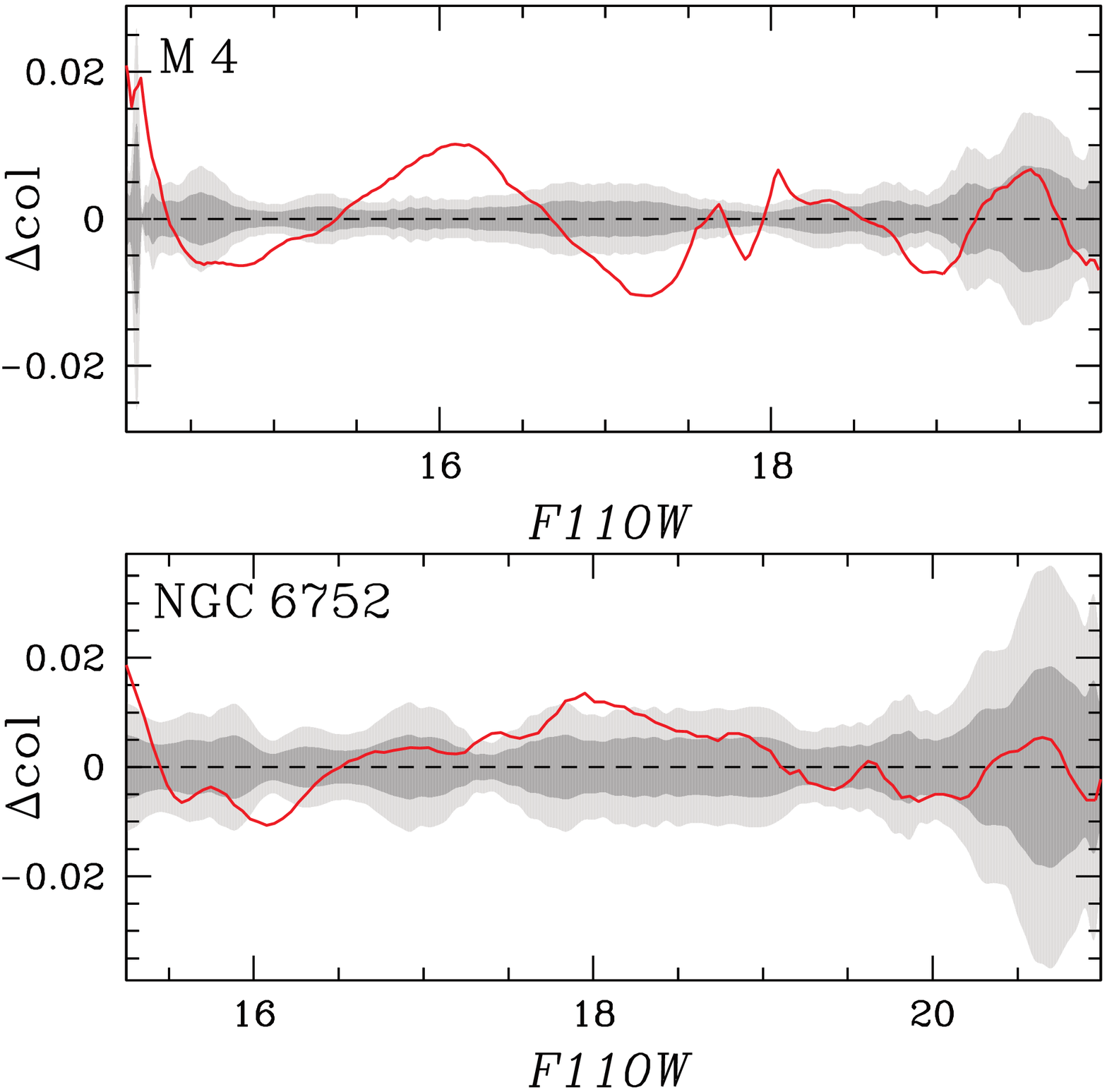}
}
\caption{Residuals of the fit between the best-fit isochrone and the fiducial line for the four clusters (Top left panel: 47\,Tuc; top right panel: M\,4; bottom left panel: NGC\,2808; bottom right panel: NGC\,6752). Dark and light gray areas represent the 1$\sigma$ and 2$\sigma$ fiducial color errors respectively.}
\label{f:residual}
\end{figure*}

Fig.~\ref{f:cmd_iso} shows the GC CMDs (ordered as in Fig.~\ref{f:cmd}) with superimposed fiducial lines (black lines) and the best-fit isochrones (red lines). The derived parameters and uncertainties from the best-fit isochrone for each cluster (i.e., age, metallicity, distance modulus, and reddening -- as described in Sect.~\ref{s:prob}) are reported in Table~\ref{t:GC_param}.  The best-fit isochrones reproduce almost perfectly the fiducial lines along all of the sequences in the CMDs. To better visualize the goodness of the fit, we illustrate the fit residuals (i.e., the term $\Delta$col in Eq.~\ref{e:chi}) in Fig.~\ref{f:residual} as a function of the magnitude F110W for the four GCs (47\,Tuc, top left panel, M\,4, top right panel, NGC\,2808, bottom left panel, and NGC\,6752, bottom right panel). The 1$\sigma$ and 2$\sigma$ errors (dark and light gray area in Fig.~\ref{f:residual}), where $\sigma$ is the fiducial color error as defined above, are also presented in Fig.~\ref{f:residual}. Small discrepancies between the fiducial line and the best-fit isochrone, of the order of few hundredths of a magnitude, are observed in the sub giant branch and red giant branch of 47\,Tuc and in the red giant branch of NGC\,6752. However, in the latter case, we note that due to the paucity of stars populating this region of the CMD, the fiducial line colors can not be determined with high precision.
For M\,4, we derived the best-fit parameters using also the other two fiducial lines obtained with the different magnitude cuts, as described in Sect.~\ref{s:cmd}. We obtained variations of just $\pm$ 0.2 Gyr for the age, $\pm$ 0.02 dex for the metallicity, $\pm$ 0.01 mag for the distance modulus, and $\pm$ 0.01 mag for the reddening, with respect to the value reported in Table~\ref{t:GC_param}. 

\begin{table*}[thb]
\begin{center}
\caption{GC parameter estimates}
\begin{tabular}{c|cc|cc|cc|cc}
\hline
\hline
Cluster & Age & $\sigma$ & [Fe/H] & $\sigma$ & $(m-M)_0$ & $\sigma$ & E\,(B-V) & $\sigma$ \\
 (1) & (2) & (3) & (4) & (5) & (6) & (7) & (8) & (9)\\
\hline
47\,Tuc   & 11.6  & $^{+0.7}_{-0.7}$ & -0.69 & $^{+0.8}_{-0.8}$ & 13.31 & $^{+0.04}_{-0.05}$ & 0.04 &  $^{+0.01}_{-0.02}$\\
 & & & & & & & & \\
M\,4      & 11.5  &  $^{+0.5}_{-0.5}$ & -1.09 & $^{+0.06}_{-0.04}$& 11.35 &  $^{+0.03}_{-0.04}$ & 0.38 & $^{+0.01}_{-0.02}$ \\
 & & & & & & & & \\
NGC\,2808 & 11.2  & $^{+1.1}_{-0.7}$  & -1.23 & $^{+0.07}_{-0.11}$ & 15.09 & $^{+0.04}_{-0.06}$ & 0.21 & $^{+0.02}_{-0.03}$ \\
 & & & & & & & & \\
NGC\,6752 & 12.1  &  $^{+1.0}_{-1.2}$ & -1.50 & $^{+0.08}_{-0.13}$ & 13.11 & $^{+0.06}_{-0.06}$ & 0.02 & $^{+0.03}_{-0.02}$ \\
 & & & & & & & & \\
\hline
\end{tabular}
\tablecomments{Columns (1): Name of the clusters. (2): Age (Gyr). (3):  Age uncertainty (Gyr, 68\% confidence interval). (4): Metallicity (dex). (5): Metallicity uncertainty (dex, 68\% confidence interval). (6) Distance modulus (mag). (7) Distance modulus uncertainty (mag, 68\% confidence interval). (8): Reddening (mag). (9) Reddening uncertainty (mag, 68\% confidence interval).}
\label{t:GC_param}
\end{center}
\end{table*} 

\begin{figure*}[thb]
\begin{center}
\includegraphics[scale=0.55]{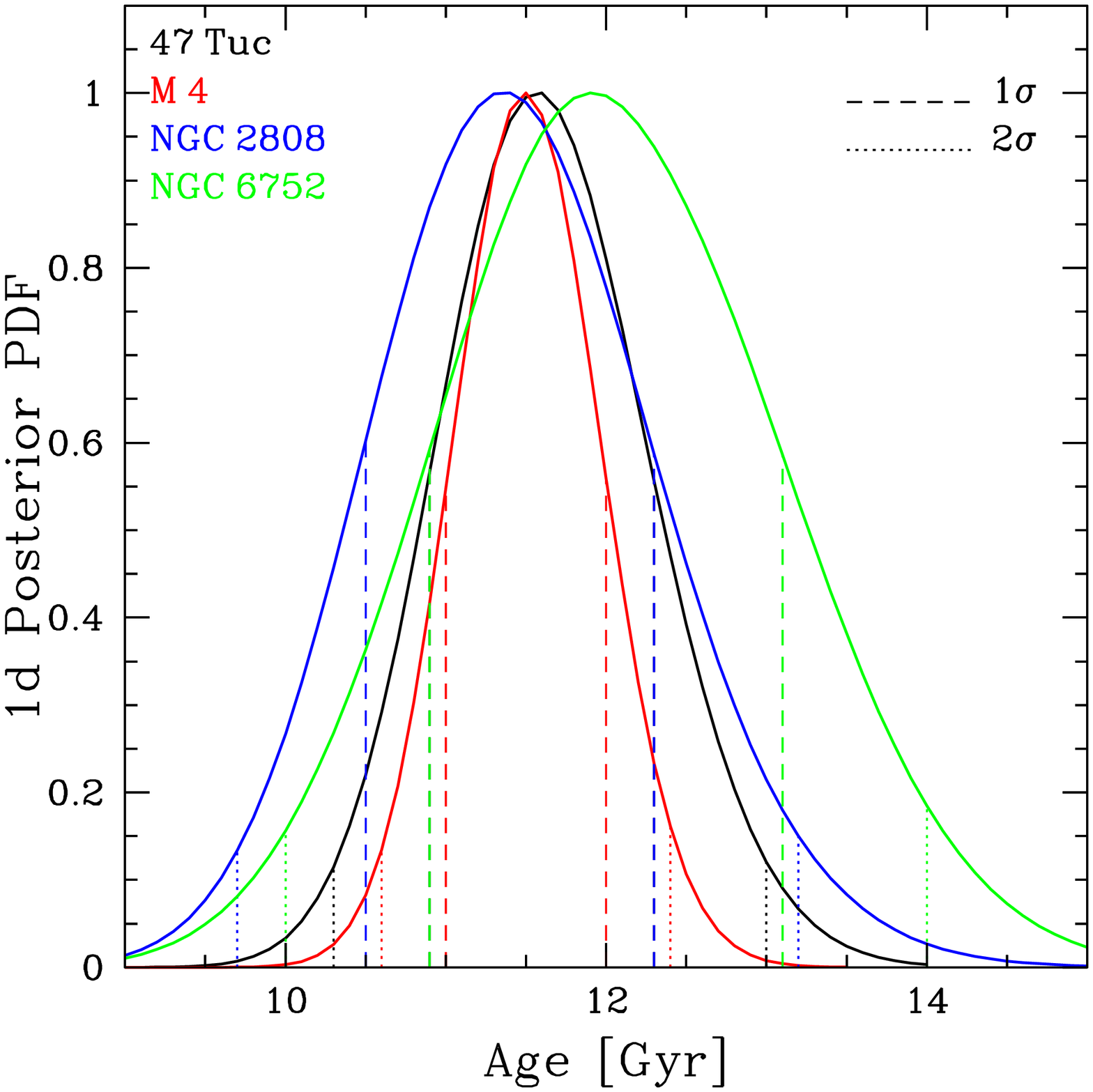}
\caption{1d posterior PDFs of the age for the four GCs (Black line: 47\,Tuc; red line: M\,4; blue line: NGC\,2808; green line: NGC\,6752) obtained by marginalizing the 4d PDFs over metallicity, distance and reddening. 1$\sigma$ (dashed lines) and 2$\sigma$ (dotted lines) confidence intervals, defined as described in Sect.~\ref{s:prob}, are also shown.}
\label{f:prob1D}
\end{center}
\end{figure*}

\subsection{Comparison with literature estimates}
\label{s:iso}
To verify that the isochrone fitting provides reasonable values for the best-fit
parameters, we compared our results with literature estimates of these four GCs.
The most direct comparison is with the results obtained by \citet{vand+14}, but,
unfortunately, there is only one cluster in common between the two studies,
namely 47\,Tuc. Hence, we focus on the results obtained by \citet{vand+13}, who
derived the same parameters from the analysis of HST Advanced Camera for Survey
(ACS) visible light photometry. This analysis used the same stellar evolutionary
code that we use, although slightly different prescriptions for the isochrone
computation. Briefly, the main difference between the two sets of models
concerns the assumed metal mixture, that is \citet{vand+14} models assume a
significantly lower abundance of the CNO elements \citep[see for a detailed
description of the models and their differences,][]{vand+13,vand+14}. This,
in turn, implies that slightly higher ages will be obtained using
\citet{vand+14} models, with respect to those derived by \citet{vand+13}, due to
the fact that turn-off luminosity versus age relations depend on the absolute
abundance of oxygen \citep[see for an exhaustive discussion,][]{vand+12}. In
this context, \citet{vand+14} specified that to partially compensate for the
expected effects of the different oxygen and metal abundances between the
models, they assumed slightly larger distance moduli ($\leq 0.05$ mag) and ages
(by 0.25 Gyr) with respect to the values derived by \citet{vand+13}. We can
then assume that a similar variation, of the order of 0.25\,--\,0.75 Gyr
(depending on the assumed apparent distance modulus), can be expected also for
the clusters in this work that were not analyzed in \citet{vand+14}; therefore,
the ages reported in \citet{vand+13} should be increased by these quantities in
order to obtain ages based on \citet{vand+14} isochrones. We note that, also
taking into account these variations, our age estimates are in agreement within
the errors with those derived by \citet{vand+13}. In the following, to not
generate confusion to the readers, we will report the original values derived by
\citet{vand+13}, with the exception of 47\,Tuc for which we report the values
derived in \citet{vand+14}.

We briefly summarize these comparisons.

{\it 47\,Tuc}: the best-fit parameters for this cluster are in good agreement within the errors with the values derived by \citet{vand+14}. They derived an age of 12.0 Gyr and an apparent distance modulus $(m-M)_V$ = 13.40 mag, assuming a metallicity [Fe/H] = $-$0.76 dex and reddening E(B-V) = 0.028 mag\footnote{\citet{vand+13} and \citet{vand+14} adopted [Fe/H] values from \citet{carr+09b} and E(B-V) values, with few exceptions, from \citet{schl+98}.}.
Our best-fit parameters are also in good agreement with the \citet{harr06} 
database; [Fe/H] = $-$0.72 dex, $(m-M)_V$ = 13.37 mag and E(B-V) = 0.04 mag.

{\it M\,4}: the derived values for M\,4 are in good agreement with the values
derived by \citet{vand+13}, with the exception of the metallicity. 

Within the uncertainty, we derive the same age (11.5 Gyr) and apparent distance
modulus  ($(m-M)_V = 12.72$ mag vs $(m-M)_V = 12.74$ mag) as \citet{vand+13},
but the metallicity  is slightly different ([Fe/H] = -1.09 dex vs [Fe/H] =
$-$1.18 dex). However, the metallicity  estimates of M\,4 derived from
high-resolution spectroscopy cover this range; for example,  [Fe/H] $\simeq$
$-$1.07 dex \citep{mari+08,vill+12,mala+13} to [Fe/H] $\simeq$ $-$1.18 
\citep{ivan+99,carr+09a}. As shown in Table~\ref{t:GC_prior}, we adopted [Fe/H]
= $-$1.12  dex \citep{mari+11} as a reference value. 

The M\,4 dataset in our study is optimized to achieve high-precision photometry
for  faint M-dwarfs.  As a result, we were not able to sample the RGB and could
not  achieve a stronger constraint on the cluster metallicity from CMD fitting.
Still, the derived metallicity is in good agreement with the literature values, 
suggesting that the luminosity of the kink and the bending of the low-mass MS 
could provide a new diagnostic for this parameter. Distance determination also 
agrees quite well with the recent estimates based on RR Lyraes \citep[$(m-M)_0$
=  11.27\,--\,11.39 mag,][see their Table~1 for a detailed summary of the
distance  and reddening available in the literature]{neel+15,brag+15}. Taking
into account the derived reddening E(B-V) = 0.38 mag and the adopted $R_V$ =
3.6, we obtain an apparent distance modulus $(m-M)_V$ = 12.72 mag. Since $<M_V>$
= 13.33 mag for the RR Lyrae in M\,4 \citep{brag+15}, we obtain a $<M_V>$ = 0.61
mag. \citet{clem+03} results for the apparent magnitude of the Large Magellanic
Clouds (LMC) RR Lyrae as a function of [Fe/H] coupled with the very precise LMC
distance found by \citet{piet+13} yields $<M_V>$ = 0.66 $\pm$ 0.06 mag for RR Lyrae that have [Fe/H] values near $-$1.1 dex, in agreement within the errors
with the $<M_V>$ derived in this work.  

{\it NGC\,2808}: While the best-fit parameters for age, metallicity, and
reddening  are in good agreements with \citet{vand+13}, the apparent distance
modulus does not agree. Our value is $(m-M)_V$ = 15.74 mag, 0.15 mag larger
than the \citet{vand+13}  estimate of $(m-M)_V$ = 15.59 mag \citep[coincident
with the value tabulated in][]{harr06}.  Using a different stellar model grid
\citep[BaSTI,][]{piet+04,piet+06}  and a slightly different chemical composition
([$\alpha/Fe]$ = +0.3 dex), \citet{mari+14}  fitted the cluster using an
isochrone with age 10~Gyr, metallicity [Fe/H] = $-$1.15 dex,  apparent distance
modulus $(m-M)_V$ = 15.67 mag, and reddening E(B-V) = 0.19 mag.  This latter
value for the distance is in agreement within the errors with our estimate,
although from the distance-age relation we should expect to obtain a younger age
given our larger distance estimate. As another example, \citet{dale+11}, used
high-resolution far-UV and  optical images of the central region of the cluster
from HST/WFPC2 to analyze the  horizontal branch of the cluster.  Using the same
set of isochrones as in \citet{mari+14}, they derived an age of 12~Gyr, a
metallicity [Fe/H] = $-$1.31 dex, and an $[\alpha/Fe]$ =  $+$0.4 dex.  They
measured an apparent distance modulus $(m-M)_V$ = 15.74 mag, the same as that
derived in our study. 

{\it NGC\,6752}: best-fit parameters are in agreement within the uncertainties with the results obtained by \citet{vand+13}, who derived an age of 12.5 Gyr and apparent distance modulus $(m-M)_V$ = 13.24 mag, assuming a  metallicity [Fe/H] = $-$1.55 dex and reddening E(B-V) = 0.056 mag. Our distance value is also in good agreement with the value tabulated in \citet{harr06} database, $(m-M)_V$ = 13.13 mag.

\subsection{Posterior probability density function:\\ deriving age uncertainties}
\label{s:prob}

\begin{figure*}[thp]
\hspace{-0.1cm}
\includegraphics[scale=0.65,angle=270]{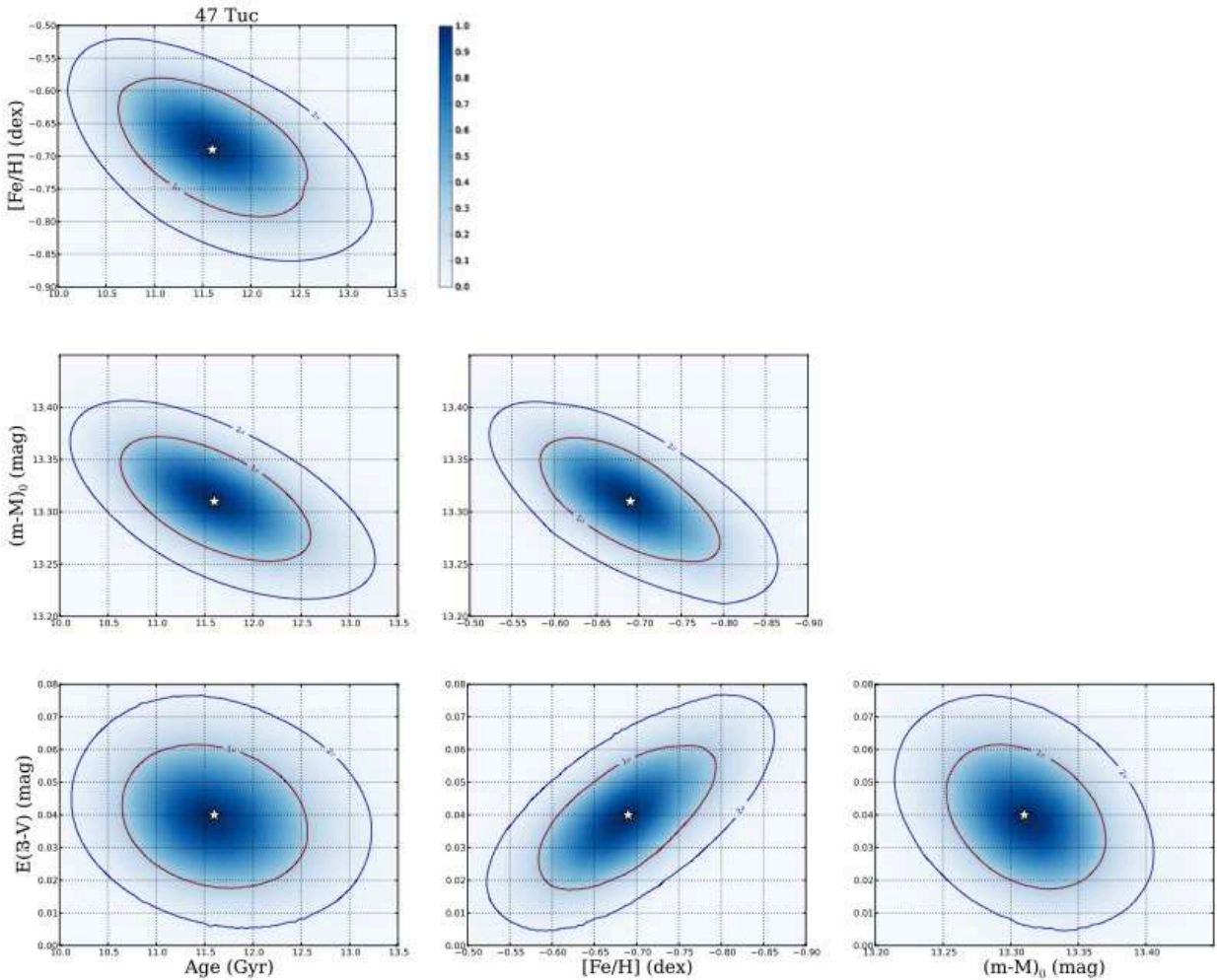}
\caption{2d posterior PDFs of all of the parameter combinations for 47\,Tuc (upper panel: age vs metallicity 2d PDF; middle panels: age vs distance modulus, metallicity vs distance modulus 2d PDFs; lower panels: age vs reddening, metallicity vs reddening, and distance modulus vs reddening 2d PDFs). The 1$\sigma$ (red lines) and 2$\sigma$ (blue lines) regions are defined as the smallest regions such that the integral of the 2d PDFs within the regions are equal to 0.68 and 0.95. Color codes for the 2d PDFs are shown in the upper right sub-panel. The white stars indicate the values for which the 4d PDF has a maximum.}
\label{f:2d_47Tuc}
\end{figure*}

\begin{figure*}[thp]
\hspace{-0.1cm}
\includegraphics[scale=0.65,angle=270]{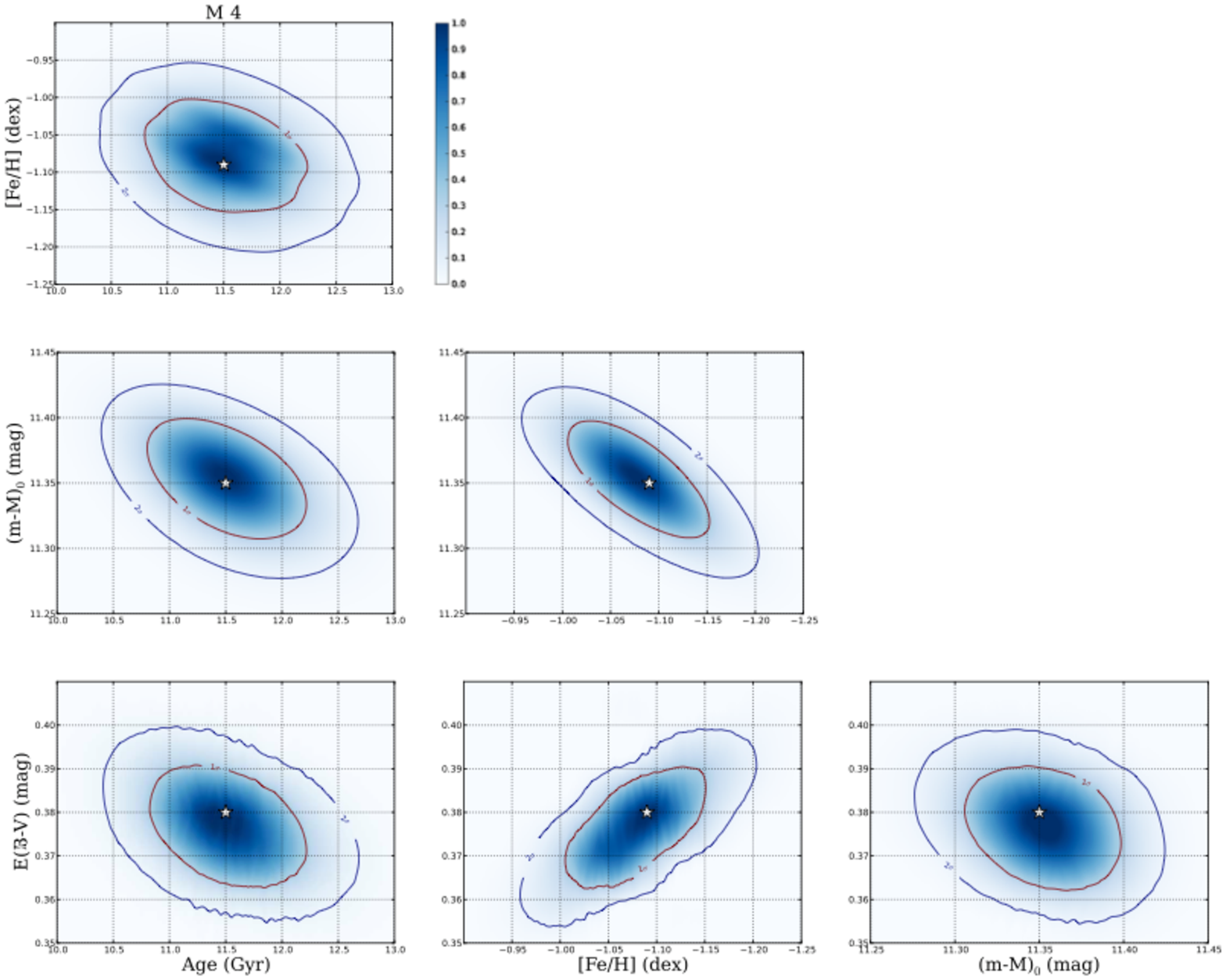}
\caption{Same as Fig.~\ref{f:2d_47Tuc}, for the GC M\,4}
\label{f:2d_ngc6121}
\end{figure*}

As we stated in Sect.~\ref{s:intro}, the main advantage of using the MS kink in IR CMDs as an age diagnostic is that this feature can provide a fundamental tool to reduce the uncertainties associated with GC age estimates. The location of the kink and the shape of the bending of the low-mass MS are dependent on metallicity, but are independent of age, allowing a partial break of the degeneracy between these two parameters. Due to the particular shape of the color-magnitude relation below the kink, the uncertainties related to the distance and reddening are also reduced.

To test the validity of this hypothesis and accurately derive the age uncertainties for each cluster, we adopted the following approach. For each isochrone of the grid, we derived the joint posterior PDF, which, as stated in Sect.~\ref{s:iso}, due to the choice of uniform priors for our parameters, is proportional to the likelihood $\mathcal{L}$. The latter is derived from Eq.~\ref{e:like} and Eq.~\ref{e:chi}. We then derived the 1d posterior PDF for the age, obtained by the marginalization of the 4d PDF (i.e., $P(age, metallicity, reddening, distance)$) over metallicity, distance and reddening. The 1d posterior PDFs for the four GCs are shown in Fig.~\ref{f:prob1D} (47\,Tuc, black line, M\,4, red line, NGC\,2808, blue line, and NGC\,6752, green line, respectively). The age uncertainties are derived from the cumulative distribution of the marginalized age distribution; 1$\sigma$ (2$\sigma$) confidence intervals are defined as the area enclosed within 16\% (2.5\%) and 84\% (97.5\%) of each cluster's cumulative distribution. This would correspond to the {\it true} 1$\sigma$ and 2$\sigma$ uncertainties if the 1d PDFs were Gaussian. Fig.~\ref{f:prob1D} shows that the derived random age uncertainties are of the order of $\sigma \sim$ 0.7\,-\,1.1 Gyr. 
As stated in Sect.~\ref{s:cmd} and considering the uncertainties in the best-fit parameters for M\,4 when a different magnitude cut is applied, we acknowledge that the derived uncertainty for the age of this cluster, $\sigma$ = 0.5 Gyr, could be underestimated. Taking into account the age range derived in Sect.~\ref{s:iso}, and considering a systematic uncertainty of the order of $\pm$ 0.3 Gyr, we estimate that the value $\sigma\sim$ 0.8 Gyr is more representative of the GC age uncertainty. Similar uncertainties, of the order of $\pm$ 1 Gyr, have been recently derived by \citet{mone+15}, which exploited the MS kink as a diagnostic to derive the absolute age of the old metal-poor GC M\,15.  Their analysis is based on near-IR images collected with the PISCES camera, coupled with the First Light Adaptive Optics system mounted at the Large Binocular Telescope.

\begin{figure*}[thp]
\hspace{-0.1cm}
\includegraphics[scale=0.65,angle=270]{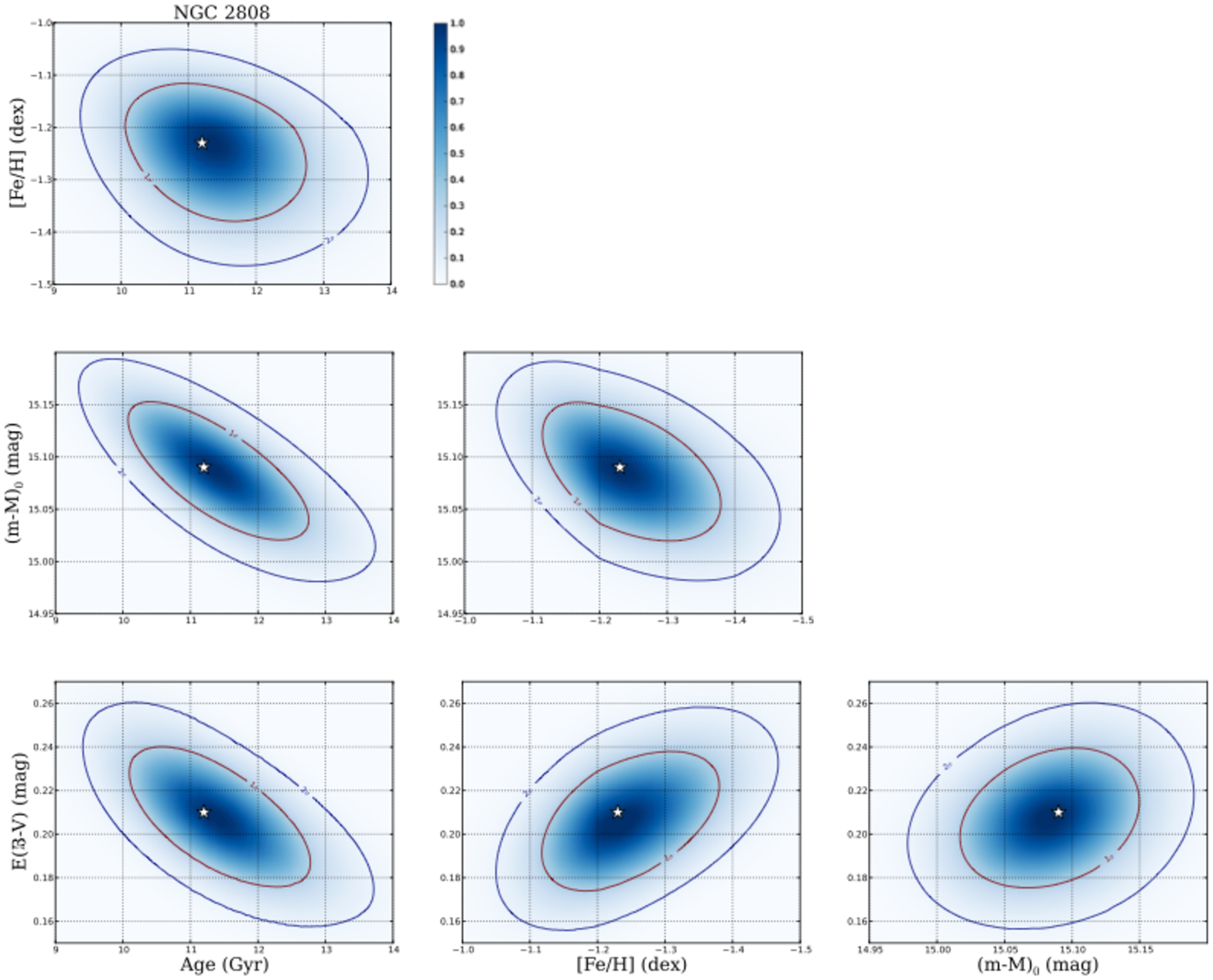}
\caption{Same as Fig.~\ref{f:2d_47Tuc}, for the GC NGC\,2808}
\label{f:2d_ngc2808}
\end{figure*}

\begin{figure*}[thp]
\hspace{-0.1cm}
\includegraphics[scale=0.65,angle=270]{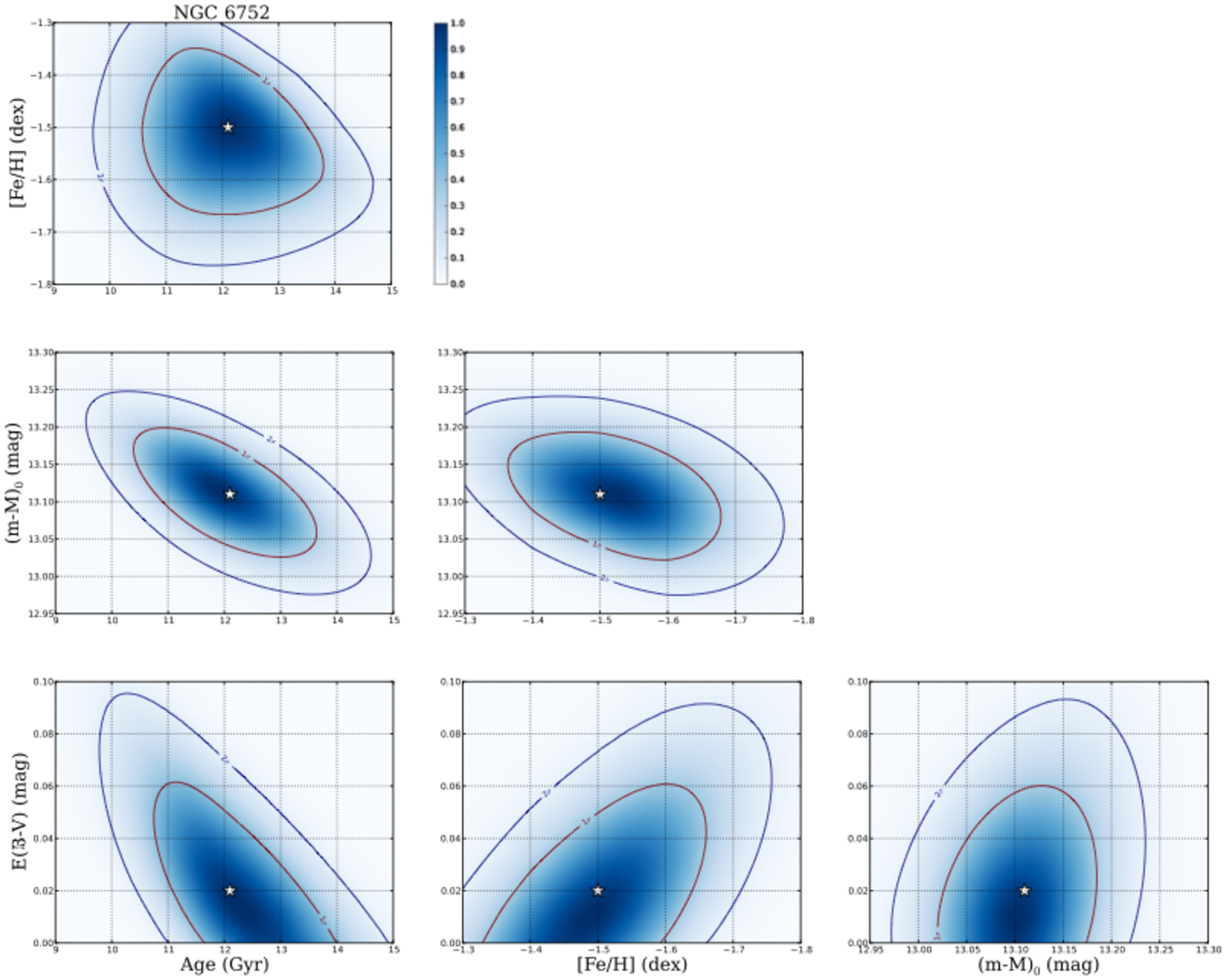}
\caption{Same as Fig.~\ref{f:2d_47Tuc}, for the GC NGC\,6752}
\label{f:2d_ngc6752}
\end{figure*}

To better visualize how the use of the kink lessens the correlations and interdependencies between the various parameters, we derived the 2d posterior PDFs of all of the parameters for each cluster. These 2d PDFs are obtained as described above, with the only difference being that the 4d PDF is marginalized over two parameters instead of three. For example, the 2d PDF of age and metallicity is obtained by marginalization of the 4d PDF over distance and reddening. Fig.~\ref{f:2d_47Tuc} shows 47\,Tuc 2d posterior PDFs, 
Fig.~\ref{f:2d_ngc6121} M\,4 2d posterior PDFs, Fig.~\ref{f:2d_ngc2808} NGC\,2808 2d posterior PDFs, and Fig.~\ref{f:2d_ngc6752} NGC\,6752 2d PDFs. In particular, the distance vs reddening 2d PDFs (bottom right panel in each Figure) shows that the morphology of the color-magnitude relation in the low mass-MS leads to a decrease in the uncertainties on distance and reddening, and consequently on the GC ages.

These uncertainties can be considered to be an upper limit as they are obtained adopting a conservative approach that assumes a priori uniform distributions over the allowed parameter ranges. If we adopted stronger constraints from independent measures of these parameters, the derived age uncertainty would be lower. Finally, we note that the quoted uncertainties only take into account random noise (number statistics and measurement errors). We acknowledge that the total uncertainty on GC parameters is further increased by the presence of a systematic component due to e.g. the choice of the stellar evolution library, stellar atmospheres and possible zero-point offsets (as well as other, more subtle, possibly unknown sources of uncertainty).

The results from this project suggest that using IR photometry and the kink as an age diagnostic can push the absolute ages of GCs to sub-Gyr accuracy with the next generation near-IR telescopes, such as the {\it James Webb Space Telescope} (JWST), the {\it Wide Field Infrared Survey Telescope} (WFIRST), and large ground-based telescopes with advanced adaptive optics technology.

\section{Summary and Conclusions}

In this study, we investigated the predictive power and the constraints established on GC properties by leveraging the shape of the MS color-magnitude relation in pure IR CMDs. The IR color-magnitude relation exhibits a kink due to opacity effects in M dwarfs, such that low mass MS stars become bluer in the IR color baseline and not redder. The combination of the MS turn-off location with that of the MS kink, and the shape of the bending below the kink, decreases the effects of the correlations and interdependencies between various parameters (i.e., age, metallicity, distance and reddening). Therefore, this diagnostic offers a new opportunity to improve the accuracy of GC age measurements over visible-light studies.

We analyzed publicly available MAST data for four GCs, namely 47\,Tuc, M\,4, NGC\,2808, and NGC\,6752, for which existing data are deep enough to reach $\simeq$ 2 mag below the kink.  With this photometry, we can fully sample 
the bending of the MS at low masses. 

By using an ad-hoc fitting method, we derived fiducial lines for the four GCs. We compared them with a grid of isochrones over a large range of parameter space that allows age, metallicity, distance, and reddening to vary within reasonably selected ranges. We calculated the joint posterior PDF for the four parameters, and derived best fit isochrones by maximizing the PDF. We obtained an age of 11.6 Gyr for 47\,Tuc, 11.5 Gyr for M\,4, 11.2 Gyr for NGC\,2808, and 12.1 Gyr for NGC\,6752. These best-fit ages, and the other derived parameters, agree quite well with the results obtained by \citet{vand+13}, who used the same stellar evolution code to analyze visible-light ACS photometry of these GCs.  

To derive the age uncertainty in each GC, we calculated the 1d posterior PDF, obtained by marginalizing the 4d posterior PDF over metallicity, distance and reddening. We calculated the 1$\sigma$ and 2$\sigma$ confidence intervals from the cumulative distribution of the marginalized 1d PDF, where 1(2)$\sigma$ is defined as the area enclosed within the 16\% (2.5\%) and 84\% (97.5\%) of the cumulative distribution. The random age uncertainties are $\sigma$ $\sim$ 0.7\,--\,1.1 Gyr. Our results suggest that the IR color-magnitude relation and kink in the lower MS represents a promising tool to obtain absolute ages of GCs with sub-Gyr accuracy.
   
Deriving sub-Gyr ages for GCs represents a foundation for many astrophysical topics. A precise age will constrain the formation epoch of clusters with respect to reionization, which is crucial for constraining models of cluster formation.  More accurate ages will also lead to a new age-metallicity relation with absolute uncertainty that is smaller than the current scatter. Obtaining sub-Gyr ages for GCs will enable us to accurately derive the {\it absolute} normalization and the slope of this relation. The latter is an anchor to high-resolution N-body simulations of galaxy formation, and informs the mass-merger history that leads to the build up of Milky Way-type halos \citep[][and references therein]{macgil04,beer+12}.

Producing an accurate set of fiducial lines, spanning a large range of metallicity, is of fundamental importance both from the theoretical and observational point of view. These relations can help test stellar models for low mass stars and represent fundamental templates for the calibration of future space observations with the next generation near-IR telescopes, such as JWST, WFIRST, and ground-based 30-meters telescopes with adaptive optics technology. 

We are grateful for a very thoughtful review by the anonymous referee, which improved this paper.

\newpage
\appendix
\section{Fiducial line tables}

In the following table, we report the colors (F110W - F160W), magnitudes (F160W) and errors for the fiducial lines of each GC.

\begin{center}
\begin{longtable}{ccc|ccc|ccc|ccc}
\caption{Fiducial lines for the four GCs}\\
\hline
\multicolumn{3}{c|}{47\,Tuc} & \multicolumn{3}{c|}{M\,4} & \multicolumn{3}{c|}{NGC\,2808} & \multicolumn{3}{c}{NGC\,6752}\\
\hline
\endfirsthead
\multicolumn{12}{c}{{Continued from previous page}}\\ \hline \multicolumn{3}{c|}{47\,Tuc} & \multicolumn{3}{c|}{M\,4} & \multicolumn{3}{c|}{NGC\,2808} & \multicolumn{3}{c}{NGC\,6752}\\
\endhead
\hline \multicolumn{12}{c}{{Continued on next page}}\\
\endfoot
\hline \hline
\endlastfoot
      &       &       &       &       &       &       &      & & & & \\
F110W - F160W & F160W & $\sigma$ & F110W - F160W & F160W& $\sigma$ & F110W - F160W & F160W & $\sigma$ & F110W - F160W & F160W & $\sigma$ \\
      &       &       &       &       &       &       &      & & & & \\
0.5915 & 14.1755 & 0.0089 & 0.6265 & 13.4576 & 0.0031 & 0.7744 & 16.0513 & 0.0111 & 0.4769 & 14.4001 & 0.0037 \\
0.5905 & 14.2255 & 0.0096 & 0.6241 & 13.5076 & 0.0041 & 0.5858 & 16.1513 & 0.0047 & 0.4775 & 14.4501 & 0.0027 \\
0.5776 & 14.2755 & 0.0067 & 0.6118 & 13.5576 & 0.012 & 0.5858 & 16.2013 & 0.0053 & 0.4772 & 14.5001 & 0.0035 \\
0.575 & 14.3256 & 0.0056 & 0.604 & 13.6076 & 0.002 & 0.6954 & 16.2513 & 0.0204 & 0.4741 & 14.5501 & 0.0052 \\
0.5723 & 14.3756 & 0.0046 & 0.5984 & 13.6576 & 0.0034 & 0.5831 & 16.3513 & 0.005 & 0.4716 & 14.6001 & 0.005 \\
0.5711 & 14.4256 & 0.0037 & 0.5907 & 13.7076 & 0.0024 & 0.5799 & 16.4013 & 0.005 & 0.469 & 14.6501 & 0.0052 \\
0.5706 & 14.4756 & 0.0033 & 0.5852 & 13.7576 & 0.0024 & 0.5777 & 16.4513 & 0.0052 & 0.4674 & 14.7001 & 0.0059 \\
0.5694 & 14.5256 & 0.003 & 0.5796 & 13.8076 & 0.0024 & 0.6041 & 16.5013 & 0.0094 & 0.4647 & 14.7501 & 0.0063 \\
0.568 & 14.5756 & 0.0028 & 0.5744 & 13.8576 & 0.0024 & 0.5726 & 16.5513 & 0.0054 & 0.463 & 14.8001 & 0.0058 \\
0.5663 & 14.6256 & 0.0027 & 0.5696 & 13.9076 & 0.0031 & 0.5713 & 16.6013 & 0.0048 & 0.4601 & 14.8501 & 0.0054 \\
0.5643 & 14.6756 & 0.0027 & 0.565 & 13.9576 & 0.0034 & 0.5684 & 16.6513 & 0.0047 & 0.456 & 14.9001 & 0.0047 \\
0.5625 & 14.7256 & 0.0029 & 0.5606 & 14.0076 & 0.0035 & 0.5653 & 16.7013 & 0.0046 & 0.4513 & 14.9501 & 0.0043 \\
0.5611 & 14.7756 & 0.0032 & 0.5564 & 14.0576 & 0.0032 & 0.5622 & 16.7513 & 0.0045 & 0.4454 & 15.0001 & 0.0041 \\
0.5594 & 14.8256 & 0.0033 & 0.553 & 14.1076 & 0.0027 & 0.5603 & 16.8013 & 0.0043 & 0.4392 & 15.0501 & 0.0042 \\
0.5572 & 14.8756 & 0.0031 & 0.55 & 14.1576 & 0.0023 & 0.5573 & 16.8513 & 0.0039 & 0.4322 & 15.1001 & 0.0044 \\
0.5542 & 14.9256 & 0.0031 & 0.5477 & 14.2076 & 0.0017 & 0.554 & 16.9013 & 0.0036 & 0.4251 & 15.1501 & 0.0047 \\
0.5509 & 14.9756 & 0.003 & 0.5457 & 14.2576 & 0.0015 & 0.5496 & 16.9513 & 0.0035 & 0.4174 & 15.2001 & 0.005 \\
0.5479 & 15.0256 & 0.0029 & 0.5438 & 14.3076 & 0.0012 & 0.5441 & 17.0013 & 0.0034 & 0.4098 & 15.2501 & 0.0049 \\
0.5458 & 15.0756 & 0.0028 & 0.5421 & 14.3576 & 0.0011 & 0.5364 & 17.0513 & 0.0038 & 0.4028 & 15.3001 & 0.0048 \\
0.5429 & 15.1256 & 0.0026 & 0.5406 & 14.4076 & 0.001 & 0.5262 & 17.1013 & 0.0039 & 0.3965 & 15.3501 & 0.0046 \\
0.5394 & 15.1756 & 0.0027 & 0.5393 & 14.4576 & 0.001 & 0.5158 & 17.1513 & 0.0038 & 0.3917 & 15.4001 & 0.0052 \\
0.5346 & 15.2256 & 0.0028 & 0.5386 & 14.5076 & 0.001 & 0.5057 & 17.2013 & 0.0034 & 0.3876 & 15.4501 & 0.0055 \\
0.5278 & 15.2756 & 0.0034 & 0.5383 & 14.5576 & 0.0012 & 0.4968 & 17.2513 & 0.0035 & 0.3848 & 15.5001 & 0.0058 \\
0.5207 & 15.3256 & 0.0035 & 0.5388 & 14.6076 & 0.0014 & 0.4897 & 17.3013 & 0.0036 & 0.3819 & 15.5501 & 0.0053 \\
0.5137 & 15.3756 & 0.0032 & 0.5394 & 14.6576 & 0.0016 & 0.4826 & 17.3513 & 0.0034 & 0.3798 & 15.6001 & 0.0044 \\
0.5048 & 15.4256 & 0.0029 & 0.5403 & 14.7076 & 0.0017 & 0.4763 & 17.4013 & 0.0031 & 0.3774 & 15.6501 & 0.0033 \\
0.495 & 15.4756 & 0.0036 & 0.5409 & 14.7576 & 0.0017 & 0.4717 & 17.4513 & 0.0026 & 0.3754 & 15.7001 & 0.0026 \\
0.4845 & 15.5256 & 0.0039 & 0.5418 & 14.8076 & 0.0017 & 0.4676 & 17.5013 & 0.0024 & 0.373 & 15.7501 & 0.0024 \\
0.4725 & 15.5756 & 0.0037 & 0.5426 & 14.8576 & 0.0014 & 0.4635 & 17.5513 & 0.0024 & 0.3705 & 15.8001 & 0.0028 \\
0.4612 & 15.6256 & 0.0034 & 0.5437 & 14.9076 & 0.0013 & 0.4589 & 17.6013 & 0.0028 & 0.3681 & 15.8501 & 0.0032 \\
0.4523 & 15.6756 & 0.0033 & 0.5453 & 14.9576 & 0.0012 & 0.4536 & 17.6513 & 0.0028 & 0.3656 & 15.9001 & 0.0036 \\
0.4436 & 15.7256 & 0.0029 & 0.5472 & 15.0076 & 0.0011 & 0.4486 & 17.7013 & 0.0023 & 0.3636 & 15.9501 & 0.0037 \\
0.436 & 15.7756 & 0.0025 & 0.5493 & 15.0576 & 0.0011 & 0.4447 & 17.7513 & 0.0023 & 0.3619 & 16.0001 & 0.0036 \\
0.4304 & 15.8256 & 0.0023 & 0.5515 & 15.1076 & 0.0011 & 0.4416 & 17.8013 & 0.0023 & 0.3609 & 16.0501 & 0.0034 \\
0.4252 & 15.8756 & 0.0017 & 0.554 & 15.1576 & 0.0014 & 0.4392 & 17.8513 & 0.0024 & 0.3606 & 16.1001 & 0.0031 \\
0.4206 & 15.9256 & 0.0015 & 0.5563 & 15.2076 & 0.0012 & 0.4374 & 17.9013 & 0.0021 & 0.3608 & 16.1501 & 0.003 \\
0.4169 & 15.9756 & 0.0017 & 0.5592 & 15.2576 & 0.0016 & 0.4369 & 17.9513 & 0.0018 & 0.3611 & 16.2001 & 0.003 \\
0.4131 & 16.0256 & 0.0015 & 0.5621 & 15.3076 & 0.0014 & 0.4369 & 18.0013 & 0.0016 & 0.3623 & 16.2501 & 0.0032 \\
0.4095 & 16.0756 & 0.0015 & 0.5657 & 15.3576 & 0.0015 & 0.4371 & 18.0513 & 0.0017 & 0.3634 & 16.3001 & 0.0034 \\
0.4067 & 16.1256 & 0.0015 & 0.5692 & 15.4076 & 0.0013 & 0.4372 & 18.1013 & 0.002 & 0.365 & 16.3501 & 0.004 \\
0.4048 & 16.1756 & 0.0017 & 0.5732 & 15.4576 & 0.0013 & 0.4373 & 18.1513 & 0.0025 & 0.3676 & 16.4001 & 0.0046 \\
0.4038 & 16.2256 & 0.0018 & 0.5776 & 15.5076 & 0.0013 & 0.4376 & 18.2013 & 0.0029 & 0.3697 & 16.4501 & 0.0052 \\
0.4039 & 16.2756 & 0.0019 & 0.5826 & 15.5576 & 0.0013 & 0.4385 & 18.2513 & 0.003 & 0.3723 & 16.5001 & 0.0056 \\
0.4043 & 16.3256 & 0.0019 & 0.5884 & 15.6076 & 0.0015 & 0.4401 & 18.3013 & 0.0027 & 0.3746 & 16.5501 & 0.0059 \\
0.4053 & 16.3756 & 0.002 & 0.595 & 15.6576 & 0.0015 & 0.4424 & 18.3513 & 0.0025 & 0.3775 & 16.6001 & 0.0054 \\
0.4065 & 16.4256 & 0.002 & 0.6024 & 15.7076 & 0.0016 & 0.4446 & 18.4013 & 0.0024 & 0.3803 & 16.6501 & 0.0055 \\
0.4083 & 16.4755 & 0.0018 & 0.6105 & 15.7576 & 0.0016 & 0.4461 & 18.4513 & 0.0022 & 0.3842 & 16.7001 & 0.0053 \\
0.4106 & 16.5255 & 0.0018 & 0.6193 & 15.8076 & 0.0017 & 0.4468 & 18.5013 & 0.0022 & 0.3878 & 16.75 & 0.0048 \\
0.4126 & 16.5755 & 0.0018 & 0.6284 & 15.8576 & 0.0018 & 0.4475 & 18.5513 & 0.0019 & 0.3914 & 16.8 & 0.0038 \\
0.4145 & 16.6255 & 0.0014 & 0.6387 & 15.9076 & 0.002 & 0.4486 & 18.6013 & 0.002 & 0.3947 & 16.85 & 0.0034 \\
0.4165 & 16.6755 & 0.0013 & 0.6493 & 15.9576 & 0.002 & 0.4503 & 18.6513 & 0.0023 & 0.3968 & 16.9 & 0.0024 \\
0.4185 & 16.7255 & 0.0016 & 0.6612 & 16.0076 & 0.0024 & 0.4524 & 18.7013 & 0.0025 & 0.3996 & 16.95 & 0.0024 \\
0.4217 & 16.7755 & 0.0018 & 0.673 & 16.0576 & 0.0024 & 0.4555 & 18.7513 & 0.0026 & 0.4027 & 17.0 & 0.0022 \\
0.4256 & 16.8255 & 0.0019 & 0.6857 & 16.1076 & 0.0025 & 0.459 & 18.8013 & 0.003 & 0.4065 & 17.05 & 0.0027 \\
0.4293 & 16.8755 & 0.0018 & 0.6985 & 16.1576 & 0.0023 & 0.4628 & 18.8513 & 0.0029 & 0.4114 & 17.1 & 0.0033 \\
0.4333 & 16.9255 & 0.0015 & 0.7117 & 16.2076 & 0.0023 & 0.4671 & 18.9013 & 0.0023 & 0.4164 & 17.15 & 0.0036 \\
0.4378 & 16.9755 & 0.0016 & 0.7255 & 16.2576 & 0.0024 & 0.4708 & 18.9513 & 0.0022 & 0.4209 & 17.2 & 0.0046 \\
0.4426 & 17.0255 & 0.0021 & 0.7391 & 16.3076 & 0.0023 & 0.4745 & 19.0013 & 0.0018 & 0.4257 & 17.25 & 0.0048 \\
0.4486 & 17.0755 & 0.0022 & 0.7532 & 16.3576 & 0.0024 & 0.4785 & 19.0513 & 0.0018 & 0.4292 & 17.3 & 0.0054 \\
0.4558 & 17.1255 & 0.0025 & 0.7669 & 16.4076 & 0.0023 & 0.4822 & 19.1013 & 0.0017 & 0.4337 & 17.35 & 0.0056 \\
0.463 & 17.1755 & 0.0027 & 0.7805 & 16.4576 & 0.0024 & 0.4858 & 19.1513 & 0.0018 & 0.4374 & 17.4 & 0.0059 \\
0.471 & 17.2255 & 0.0026 & 0.7931 & 16.5076 & 0.0022 & 0.4896 & 19.2013 & 0.0019 & 0.4436 & 17.45 & 0.0054 \\
0.4796 & 17.2755 & 0.0029 & 0.8048 & 16.5576 & 0.0021 & 0.4939 & 19.2513 & 0.0024 & 0.4493 & 17.5 & 0.0048 \\
0.4874 & 17.3255 & 0.0028 & 0.8153 & 16.6076 & 0.0018 & 0.4994 & 19.3013 & 0.0026 & 0.4577 & 17.55 & 0.0048 \\
0.495 & 17.3755 & 0.0023 & 0.8241 & 16.6576 & 0.0017 & 0.5063 & 19.3513 & 0.0026 & 0.4657 & 17.6 & 0.0046 \\
0.5037 & 17.4255 & 0.0027 & 0.8314 & 16.7076 & 0.0014 & 0.5137 & 19.4013 & 0.0027 & 0.475 & 17.65 & 0.0052 \\
0.5119 & 17.4755 & 0.0032 & 0.837 & 16.7576 & 0.0014 & 0.5212 & 19.4513 & 0.0028 & 0.4837 & 17.7 & 0.0052 \\
0.5213 & 17.5255 & 0.0031 & 0.8418 & 16.8076 & 0.0013 & 0.5296 & 19.5013 & 0.0028 & 0.494 & 17.75 & 0.0055 \\
0.5326 & 17.5755 & 0.0034 & 0.8454 & 16.8576 & 0.0013 & 0.5383 & 19.5513 & 0.0031 & 0.5034 & 17.8 & 0.0052 \\
0.5437 & 17.6255 & 0.004 & 0.8485 & 16.9076 & 0.0011 & 0.5477 & 19.6013 & 0.0032 & 0.5152 & 17.85 & 0.0055 \\
0.5553 & 17.6755 & 0.0036 & 0.8509 & 16.9576 & 0.0011 & 0.5581 & 19.6513 & 0.0034 & 0.5266 & 17.9 & 0.0049 \\
0.5694 & 17.7255 & 0.0044 & 0.8527 & 17.0076 & 0.001 & 0.569 & 19.7013 & 0.0035 & 0.5393 & 17.95 & 0.0054 \\
0.5838 & 17.7755 & 0.0048 & 0.8538 & 17.0576 & 0.0009 & 0.5794 & 19.7513 & 0.0036 & 0.5506 & 18.0 & 0.0052 \\
0.5976 & 17.8255 & 0.0048 & 0.8542 & 17.1076 & 0.0009 & 0.5911 & 19.8013 & 0.0036 & 0.5628 & 18.05 & 0.0051 \\
0.6135 & 17.8755 & 0.0049 & 0.8538 & 17.1576 & 0.001 & 0.6025 & 19.8513 & 0.0037 & 0.5732 & 18.1 & 0.0052 \\
0.6288 & 17.9255 & 0.0049 & 0.8527 & 17.2076 & 0.0012 & 0.6134 & 19.9013 & 0.0041 & 0.5852 & 18.15 & 0.0054 \\
0.6422 & 17.9755 & 0.0044 & 0.8508 & 17.2576 & 0.0013 & 0.6257 & 19.9513 & 0.0039 & 0.5961 & 18.2 & 0.0052 \\
0.6557 & 18.0255 & 0.0039 & 0.8484 & 17.3076 & 0.0016 & 0.6388 & 20.0013 & 0.0043 & 0.6085 & 18.25 & 0.0053 \\
0.6688 & 18.0755 & 0.004 & 0.8458 & 17.3576 & 0.0018 & 0.6512 & 20.0513 & 0.0044 & 0.6203 & 18.3 & 0.0055 \\
0.6804 & 18.1255 & 0.004 & 0.8431 & 17.4076 & 0.002 & 0.6635 & 20.1013 & 0.0041 & 0.6323 & 18.35 & 0.0055 \\
0.6912 & 18.1755 & 0.0038 & 0.8408 & 17.4576 & 0.002 & 0.6759 & 20.1513 & 0.004 & 0.6432 & 18.4 & 0.0053 \\
0.7025 & 18.2255 & 0.0035 & 0.8387 & 17.5076 & 0.002 & 0.6868 & 20.2013 & 0.0041 & 0.6518 & 18.45 & 0.0048 \\
0.712 & 18.2755 & 0.0037 & 0.837 & 17.5576 & 0.0019 & 0.6978 & 20.2513 & 0.0043 & 0.6589 & 18.5 & 0.0041 \\
0.7204 & 18.3255 & 0.0031 & 0.8356 & 17.6076 & 0.0017 & 0.7087 & 20.3013 & 0.0043 & 0.6624 & 18.55 & 0.0034 \\
0.7286 & 18.3755 & 0.0026 & 0.8343 & 17.6576 & 0.0017 & 0.7188 & 20.3513 & 0.0037 & 0.6655 & 18.6 & 0.0033 \\
0.7344 & 18.4255 & 0.0023 & 0.8332 & 17.7076 & 0.0017 & 0.7278 & 20.4013 & 0.0031 & 0.6666 & 18.65 & 0.0032 \\
0.7385 & 18.4755 & 0.0018 & 0.8322 & 17.7576 & 0.0021 & 0.7361 & 20.4513 & 0.003 & 0.6678 & 18.7 & 0.0036 \\
0.7416 & 18.5255 & 0.0017 & 0.8314 & 17.8076 & 0.0023 & 0.7431 & 20.5013 & 0.0028 & 0.6688 & 18.75 & 0.0036 \\
0.7432 & 18.5755 & 0.0018 & 0.8306 & 17.8576 & 0.0025 & 0.7488 & 20.5513 & 0.0026 & 0.6694 & 18.8 & 0.0039 \\
0.7441 & 18.6255 & 0.0019 & 0.8299 & 17.9076 & 0.0023 & 0.7534 & 20.6013 & 0.0024 & 0.6696 & 18.85 & 0.0037 \\
0.7444 & 18.6755 & 0.0019 & 0.8288 & 17.9576 & 0.0023 & 0.7565 & 20.6513 & 0.0022 & 0.6691 & 18.9 & 0.0041 \\
0.7444 & 18.7255 & 0.002 & 0.8276 & 18.0076 & 0.0019 & 0.7583 & 20.7013 & 0.0024 & 0.6685 & 18.95 & 0.0038 \\
0.7434 & 18.7755 & 0.002 & 0.8264 & 18.0576 & 0.0019 & 0.76 & 20.7513 & 0.0027 & 0.6666 & 19.0 & 0.0044 \\
0.7416 & 18.8255 & 0.002 & 0.8255 & 18.1076 & 0.0016 & 0.76 & 20.8013 & 0.0029 & 0.6668 & 19.05 & 0.0049 \\
0.739 & 18.8755 & 0.0021 & 0.8247 & 18.1576 & 0.002 & 0.7598 & 20.8513 & 0.0033 & 0.6657 & 19.1 & 0.006 \\
0.7364 & 18.9255 & 0.0025 & 0.8241 & 18.2076 & 0.0024 & 0.7593 & 20.9013 & 0.0039 & 0.6666 & 19.15 & 0.0061 \\
0.7334 & 18.9755 & 0.0025 & 0.8221 & 18.2576 & 0.0033 & 0.7576 & 20.9513 & 0.004 & 0.6658 & 19.2 & 0.0066 \\
0.7309 & 19.0255 & 0.0027 & 0.8203 & 18.3076 & 0.0039 & 0.756 & 21.0013 & 0.0046 & 0.6661 & 19.25 & 0.0052 \\
0.7283 & 19.0755 & 0.0025 & 0.8166 & 18.3576 & 0.0049 & 0.7534 & 21.0513 & 0.0042 & 0.6648 & 19.3 & 0.0052 \\
0.7254 & 19.1255 & 0.0025 & 0.8141 & 18.4076 & 0.0048 & 0.7521 & 21.1013 & 0.0045 & 0.6636 & 19.35 & 0.0048 \\
0.7229 & 19.1755 & 0.0023 & 0.811 & 18.4576 & 0.0044 & 0.7514 & 21.1513 & 0.0052 & 0.663 & 19.4 & 0.0052 \\
0.7201 & 19.2255 & 0.0024 & 0.8095 & 18.5076 & 0.0041 & 0.7486 & 21.2013 & 0.0045 & 0.6627 & 19.45 & 0.0057 \\
0.7177 & 19.2755 & 0.0025 & 0.8075 & 18.5576 & 0.0044 & 0.7474 & 21.2513 & 0.0042 & 0.6626 & 19.5 & 0.0067 \\
0.7148 & 19.3255 & 0.003 & 0.8065 & 18.6076 & 0.0053 & 0.7462 & 21.3013 & 0.004 & 0.6614 & 19.55 & 0.0078 \\
0.7119 & 19.3755 & 0.003 & 0.8046 & 18.6576 & 0.0062 & 0.7438 & 21.3513 & 0.0032 & 0.6592 & 19.6 & 0.0096 \\
0.7088 & 19.4255 & 0.0031 & 0.8032 & 18.7076 & 0.0072 & 0.7422 & 21.4013 & 0.0033 & 0.656 & 19.65 & 0.0106 \\
0.7055 & 19.4755 & 0.003 & 0.8021 & 18.7576 & 0.0072 & 0.7398 & 21.4513 & 0.0031 & 0.6539 & 19.7 & 0.0106 \\
0.7022 & 19.5255 & 0.0036 & 0.8019 & 18.8076 & 0.007 & 0.7376 & 21.5013 & 0.003 & 0.6525 & 19.75 & 0.0107 \\
0.6989 & 19.5755 & 0.0036 & 0.8035 & 18.8576 & 0.0068 & 0.7363 & 21.5513 & 0.0026 & 0.6517 & 19.8 & 0.0113 \\
0.6956 & 19.6255 & 0.0043 & 0.8057 & 18.9076 & 0.0062 & 0.735 & 21.6013 & 0.0029 & 0.6512 & 19.85 & 0.0137 \\
0.6917 & 19.6755 & 0.0042 & 0.8074 & 18.9576 & 0.0056 & 0.7335 & 21.6513 & 0.0031 & 0.6499 & 19.9 & 0.0156 \\
0.6881 & 19.7255 & 0.0045 & 0.8095 & 19.0076 & 0.0045 & 0.7321 & 21.7013 & 0.0036 & 0.6485 & 19.95 & 0.0176 \\
0.685 & 19.7755 & 0.0041 & 0.8101 & 19.0576 & 0.004 & 0.731 & 21.7513 & 0.0032 & 0.647 & 20.0 & 0.018 \\
0.6821 & 19.8255 & 0.0045 & 0.8099 & 19.1076 & 0.005 & 0.7305 & 21.8013 & 0.0034 & 0.6473 & 20.05 & 0.0184 \\
0.6804 & 19.8755 & 0.0049 & 0.8108 & 19.1576 & 0.0051 & 0.7303 & 21.8513 & 0.0036 & 0.6495 & 20.1 & 0.0173 \\
0.6792 & 19.9255 & 0.0065 & & & & 0.7295 & 21.9013 & 0.0041 & 0.6533 & 20.15 & 0.0158 \\
0.6766 & 19.9755 & 0.0066 & & & & 0.7291 & 21.9513 & 0.0035 & 0.6559 & 20.2 & 0.0144 \\
0.6743 & 20.0255 & 0.0072 & & & & 0.7281 & 22.0013 & 0.0036 & 0.658 & 20.25 & 0.0132 \\
0.6705 & 20.0755 & 0.0069 & & & & 0.7268 & 22.0513 & 0.0036 & 0.6579 & 20.3 & 0.0153 \\
0.6675 & 20.1255 & 0.0067 & & & & 0.7258 & 22.1013 & 0.0048 & 0.6525 & 20.35 & 0.0143 \\
0.6642 & 20.1755 & 0.0072 & & & & 0.7275 & 22.1513 & 0.0048 & 0.6577 & 20.4 & 0.0328 \\
0.6623 & 20.2255 & 0.0089 & & & & 0.7271 & 22.2013 & 0.0062 & 0.6301 & 20.45 & 0.0098 \\
0.6593 & 20.2755 & 0.0106 & & & & 0.7266 & 22.2512 & 0.0075 & 0.628 & 20.5 & 0.007 \\
0.6588 & 20.3255 & 0.0116 & & & & 0.7263 & 22.3012 & 0.0068 & & & \\
0.6555 & 20.3755 & 0.0092 & & & & 0.7285 & 22.3512 & 0.0097 & & & \\
0.6547 & 20.4255 & 0.0082 & & & & 0.7298 & 22.4012 & 0.0109 & & & \\
0.6513 & 20.4755 & 0.0081 & & & & 0.7317 & 22.4512 & 0.0128 & & & \\
0.6507 & 20.5255 & 0.0083 & & & & 0.7313 & 22.5012 & 0.0122 & & & \\
0.6481 & 20.5755 & 0.0083 & & & & 0.7348 & 22.5512 & 0.0139 & & & \\
0.6464 & 20.6255 & 0.0077 & & & & 0.7356 & 22.6012 & 0.0122 & & & \\
0.6439 & 20.6755 & 0.0066 & & & & 0.7373 & 22.6512 & 0.0127 & & & \\
0.6419 & 20.7255 & 0.0063 & & & & 0.7358 & 22.7012 & 0.0122 & & & \\
0.6394 & 20.7755 & 0.0061 & & & & 0.7386 & 22.7512 & 0.011 & & & \\
0.6383 & 20.8255 & 0.0063 & & & & 0.738 & 22.8012 & 0.0125 & & & \\
0.6357 & 20.8755 & 0.0064 & & & & 0.737 & 22.8512 & 0.0137 & & & \\
0.6338 & 20.9255 & 0.0056 & & & & 0.7346 & 22.9012 & 0.0121 & & & \\
0.6318 & 20.9755 & 0.0052 & & & & 0.7399 & 22.9512 & 0.0124 & & & \\
0.631 & 21.0255 & 0.0048 & & & & 0.7347 & 23.0012 & 0.0148 & & & \\
0.6283 & 21.0755 & 0.0049 & & & & 0.7284 & 23.0512 & 0.0146 & & & \\
0.6276 & 21.1255 & 0.0053 & & & & 0.7265 & 23.1012 & 0.0102 & & & \\
0.6256 & 21.1755 & 0.0056 & & & & 0.7352 & 23.2012 & 0.0129 & & & \\
0.6245 & 21.2255 & 0.0055 & & & & 0.7157 & 23.2512 & 0.0086 & & & \\
0.623 & 21.2755 & 0.0046 & & & & 0.7133 & 23.3012 & 0.0072 & & & \\
0.6214 & 21.3255 & 0.0047 & & & & 0.7133 & 23.3512 & 0.01 &  & & \\
0.6197 & 21.3755 & 0.0044 & & & & 0.7602 & 23.4012 & 0.0288 & & & \\
0.6183 & 21.4255 & 0.0048 & & & & 0.733 & 23.4512 & 0.0397 & & & \\
0.6177 & 21.4755 & 0.005 & & & &  & & & & & \\
0.6162 & 21.5255 & 0.0051 & & & & & & & & & \\
0.6158 & 21.5755 & 0.0052 & & & & & & & & & \\
0.6143 & 21.6255 & 0.0051 & & & & & & & & & \\
0.6169 & 21.6755 & 0.0089 & & & & & & & & & \\
0.6158 & 21.7255 & 0.0116 & & & & & & & & & \\
0.6241 & 21.7755 & 0.0219 & & & & & & & & & \\
0.6276 & 21.8255 & 0.0273 & & & & & & & & & \\
0.6365 & 21.8755 & 0.0293 & & & & & & & & & \\
0.6437 & 21.9255 & 0.0264 & & & & & & & & & \\
0.6656 & 21.9755 & 0.0147 & & & & & & & & & \\
0.6428 & 22.0255 & 0.036 & & & & & & & & & \\
0.6926 & 22.0755 & 0.0427 & & & & & & & & & \\
\end{longtable}
\end{center}

\end{document}